\documentclass[useAMS,usenatbib]{mn2e}
\usepackage{graphicx}

\title[smTVS: weak LPV and NRP diagnostics]{Smoothed Temporal Variance Spectrum: weak line profile variations and NRP diagnostics}

\author[A.F.~Kholtygin and N.P.~Sudnik]{A.F.~Kholtygin$^{1}$\thanks{E-mail:
afkholtygin@gmail.com} and N.P.~Sudnik$^{1}$\\
$^{1}$Saint-Petersburg State University, Universitetskij pr. 28, Saint-Petersburg, 198504, Russia}

\begin{document}

\date{Accepted 0000 Month 00. Received 0000 Month 00; in original form 0000 Month 00}

\pagerange{\pageref{firstpage}--\pageref{lastpage}} \pubyear{0000}

\maketitle

\label{firstpage}

\begin{abstract}
We describe the version of the Temporal Variance Spectrum (TVS, \citealt{Fullerton1996}) method with pre-smoothed line profile (smoothed Temporal Variance Spectrum, smTVS). This method introduced by \citet{Kholtygin2003} can be used to detect the ultra weak variations of the line profile even for very noisy stellar spectra. 
We also describe how to estimate the mode of the non-radial pulsations (NRP) using the TVS and smTVS with different time spans. The influence of the rotational modulation of the line profile on the TVS is considered. 
The analysis of the contribution of NRP and rotational modulation in the global TVS is studied.
\end{abstract}


\section{Introduction}

Most OB stars show line profile variability (LPV). The active investigation of this phenomenon started
about 40 years ago. The LPV was found in X-ray, UV and the optical region of spectrum. It turned out that variability can be regular, cyclic and stochastic and all types of variability can present at the same time. The stochastic LPV is connected with the formation of small-scale structures in the stellar  wind. The regular and cyclic LPV can be induced by the large-scale structures in the stellar atmosphere. The presence of these structures is stimulated by wind instabilities, NRP, local inhomogeneities and the presence of global dipolar or small-scale local magnetic fields.

If the amplitude of the variations is large compared to the noise contribution and the data are well
sampled in time, it is easy to detect the LPV by overplotting individual or residual (individual profile
minus the average profile) profiles of a line or by plotting the dynamical LPV spectrum.

The commonly accepted method of detecting the relatively weak LPV is Temporal Variance Spectrum analysis introduced by~\citet{Fullerton1996}. This method can be used when the observations were obtained for both regular and irregular time grids. In this method the standard deviations of LPVs for both lines and continuum are compared. If the amplitude of deviations within a line is clearly larger than in the continuum then detection at a selected significance level can be claimed. If the observed standard deviations for LPVs are smaller than expected for the noise contribution then only the upper limit of the LPV amplitude can be estimated. 

In the following years the TVS was frequently used to study LPVs. \citet{Kaufer1996} used the TVS to investigate the long-term variations in the H$_{\alpha}$ line profiles in spectra of 6 BA supergiants.
The method of TVS was used by \cite{Markova2002} to analyse the spectral variability of O9.5\,Ia supergiant $\alpha\,$Cam. \citet{Markova2005} employed the TVS analysis to study the LPV for H$_{\alpha}$ line profiles. They  
modified the method to take into account the effects of wind emission. 

In many cases TVS analysis detects the weak LPV. For example, \citet{Oliveira2003} found variability of O\,VI $\,\lambda\,3811/34\,$\AA\ lines in spectra of WR7a star using TVS though the O\,VI lines central
depth does not exceed the noise level (see their Fig. 1). \cite{Oliveira2004} have investigated the TVS spectra for WR 48c star WX Cen. They found a highly variable component in the TVS of highly ionized ion O\,VI $\,\lambda\,3811/34\,$\AA\ lines. This component was also seen in the TVS red wings of H$_{\beta}$ and He\,II$\,\lambda 4686\,$\AA{}. The nature of the component remains unknown.

Recently \cite{Oliveira2014} have analysed the strong variability of the emission line profiles 
in spectra of the Close Binary Supersoft X-ray Source V~Sge, which is considered as a candidate to SN Ia progenitor, with TVS technique. The TVS for the strongest spectral lines were constructed using their 110 spectra obtained in 2010 and 2012. They detected only marginal LPV of the Bowen C\,III/N\,III complex ($\lambda\lambda\,4640-4650\,$\AA) and strong variations for lines He\,II $\,\lambda\,4686\,$\AA\ and H$_{\beta}$. 

\cite{Thompson2013} have presented the results of a time series analysis of 130 \'echelle spectra of $\varepsilon$ Ori (B0\,Ia), acquired over seven observing seasons between 1998 and 2006 at Ritter Observatory. TVS revealed significant wind variability in both H$_{\alpha}$ and He\,I $\,\lambda\,5876\,$\AA{} lines. 
The He\,I line TVS has a double-peaked profile consistent with radial velocity oscillations. 
In the recent paper by \cite{Martins2015} the variability of twelve spectral lines was investigated by means of the TVS for seven O9--B0.5 stars. The selected lines probe the radial structure of the atmosphere from the photosphere to the outer wind. The variability of H and He\,I--II lines was detected.

If the amplitude of LPV is sufficiently high the value of TVS within the corresponding line
substantially exceeds its value in the adjacent continuum. In many cases the LPV are extremely weak, not more than 1\% of the continuum level for many of O stars. To detect such variability the smoothed Temporal Variance Spectrum method proposed by~\citet{Kholtygin2003} can be used.

The detection of extremely weak LPV gives an opportunity to explore the stratification in the stellar atmosphere, so the detection of such micro LPV seems to be very important. We also show that the TVS (or smTVS) method can be used not only for looking for LPV.

One of the important problems in the study of non-radial pulsations of stars is determination of pulsation modes $l$ and $m$. For this purpose different methods were developed, for example~\citet{Aerts1992, Schrijvers1997, Berdyugina2003}. The most convenient of these methods is the analysis of changes of the pulsation phase component between blue and red wings of the line profile~\citep{Telting1997III}.

These methods require a large number of observations covering several periods of pulsations. It is not
always possible to carry out such observations because this would require a large number of spectra
with a high signal-to-noise ratio for each star.

Quite often collected spectra comprise only a fraction of the stellar pulsation period or the spectra are obtained at time intervals separated by large gaps. In this case using of the equations for calculation $l$ and $m$ from~\citet{Telting1997III} is impossible. However, applying TVS or smTVS is still possible.

The influence of pulsational velocity fields on photospheric line profiles was shown by \citet{Fullerton1996}. The purely radial mode produces the double-peaked TVS. When there is a mode with a large
horizontal component present, the peak separation of the TVS becomes smaller compared to a purely
radial mode. A mode with smaller horizontal velocity field produces TVS with one peak. The authors note that the shape of TVS depends on the time sampling so TVS obtained during different runs will not in general look the same.

In the present paper we describe how the TVS and smTVS method can be applied for detection of extremely small amplitude LPV in spectra of OB stars. We also studied the possibility to estimate non-radial pulsation mode using the TVS and smTVS analysis.

Our paper is organized as follow. In Sect. 2 we describe the observational material. The Smoothed Temporal Variance Spectrum is outlined in Sect. 3. Statistical properties of smTVS are described in Sect. 4. In Sect. 5 we show how to model the variable line profiles in spectra of non-radially pulsating stars. In Sects. 6 the results of our analysis together with the application to the real stellar spectra are presented in detail. The conclusions are given in Sect. 7.

\section{Observations}

Our target list consists of 3 OB supergiants.  The log of observations and stellar parameters are given in the tables~\ref{Tab.logObs} and~\ref{Tab.Par} respectively. 
\begin{table*}
 \centering
 \begin{minipage}{167mm}
  \caption{Summery of observations.}
  \label{Tab.logObs}
  \begin{tabular}{@{}lllllllll@{}}
  \hline
   HD & Name & Date & Number of & Exposition & Length of & Telescope & Spectrograph, CCD & S/N \\
      &      &      & spectra   & (min)      & observa-  &           &                   &     \\
      &      &      &           &            & tions, h  &           &                   &     \\
  \hline
   36\,486  & $\delta\,$Ori~A & 10 Jan 2004           & 40 & 3    & 2.8 & SAO, 6-m    & NES, 2k\,x\,2k      & 700 \\
   91\,316  & $\rho\,$Leo     & 10 Jan 2004           & 30 & 6    & 3.5 & SAO, 6-m    & NES, 2k\,x\,2k      & 450 \\
   210\,839 & $\lambda\,$Cep  & 20 Nov -- 16 Dec 2007 & 30 & 3--5 & 7.2 & BOAO, 1.8-m & BOES, 2048\,x\,4096 & 350 \\
  \hline
 \end{tabular}
 \end{minipage}
\end{table*}
\begin{table*}
 \centering
 \begin{minipage}{2\columnwidth}
  \caption{Stellar parameters.}
  \label{Tab.Par}
  \begin{tabular}{@{}lllcccccccrc@{}}
  \hline
   HD & Name & Spectral & V & $T_{\rm eff}$ & $\log (L/L_{\odot})$ & $M$ & $R$ & $\log g$ & $\dot{M}$ & $V\sin i$ & $V_{\infty}$ \\
   & & type & (mag) & (K) & & ($M_{\odot}$) & ($R_{\odot}$) & (cm/s$^{-2}$) & ($10^{-6}M_{\odot}/yr$) & (km/s) & (km/s) \\
  \hline 
  36\,486  & $\delta\,$Ori~A & O9.5 II        & 2.218 & 30\,600 & 5.26 & 11.2 & 11   & 3.4  & 0.76 & 157 & 2000 \\
  91\,316  & $\rho\,$Leo     & B1\,Iab        & 3.85  & 22\,000 & 5.47 & 18   & 37.4 & 2.55 & 0.35 & 75  & 1110 \\
  210\,839 & $\lambda\,$Cep  & O6.5\,I\,(n)fp & 5.043 & 36\,200 & 5.68 & 34   & 17.5 & 3.48 & 5.1  & 214 & 2200 \\
  \hline
 \end{tabular}
  
  \medskip
 The parameters for $\rho\,$Leo are from \citet{Crowther2006}. The parameters for $\lambda\,$Cep are from \citet{Markova2004}. The spectral class and $V$ for $\lambda\,$Cep and $\delta\,$Ori~A are from \citet{Sota2011} and \citet{M-A2004}, respectively. For $\delta\,$Ori~A the $T_{\rm eff}$, $\dot{M}$ and $V_{\infty}$ are from \citet{Fullerton2006}, $\log (L/L_{\odot})$, $M$, $R$, $\log g$, $V\sin i$ are from \citet{Harvin2002}.
 \end{minipage}
\end{table*}

The star $\lambda\,$Cep (HD~210839) is bright, fast rotating, runaway star, with spectral class O6.5\,I~\citep{Sota2011}. On the Hertzsprung-Russell diagram $\lambda\,$Cep is located very close to the $\beta\,$Cep stars instability strip. In 2007 the star was observed in Bohyunsan Optical Astronomy Observatory (BOAO, South Korea) by using the 1.8-m telescope and the fiber-fed echelle BOES spectrograph~\citep{Kim2002} with spectral resolution 45000 and large CCD (2048\,x\,4096 pixels). All spectra were obtained in the 3830--8260\,\AA{} region. Reduction of CCD frames was made with \textsc{iraf}. 

The supergiant $\rho\,$Leo (HD~91316) is a slowly rotating star of spectral class B1\,Iab \citep{Howarth1997}. The brightest component $\delta\,$Ori~A (HD~36486) of the wide visual triple system $\delta\,$Ori is a physical triple
system. The primary $\delta\,$Ori~Aa is an eclipsing binary (O9.5\,II and B0.5\,III~\citep{Harvin2002}) with the orbital period $P=5.73\,d$~\citep{Hoffleit1996}. The third component $\delta\,$Ori~Ab of early B subtype~\citep{Harvin2002} has the orbital period of 224.5 yr. The spectral observations for these stars were made in Special Astrophysical Observatory (SAO, Russia) on 10 January 2004 in the region $\lambda\lambda$ 4500--6000\AA{} with using 6-m telescope, the quartz echelle spectrograph NES~\citep{Panchuk2002} in Nasmyth focus with 2048\,x\,2048 Uppsala CCD. The reduction of SAO spectra was made using the \textsc{midas} package (e.g., \citealt{Kholtygin2003}). The positions of the spectral order was found by using the method by~\citet{Ballester1994}.

For studying the LPV all spectra were normalized to the individual continuum for each spectral order. 
The method of \citet{Shergin1996} was used to determine the continuum level in spectral orders with narrow spectral lines. In orders containing broad spectral lines the continuum was drawn using the procedure described by~\citet{Kholtygin2006} where polynomial approximation for all wavelengths of the order except for the regions of the broad spectral lines is used.

\section{Smoothed Temporal Variance Spectrum (smTVS). Definition and main equations}

For detection of micro LPV of various nature we used the smoothed Temporal Variance Spectrum technique, which is a modified version of the TVS analysis.

In the absence of any visible variations we can determine whether a line profile was indeed variable using a following procedure.

Before obtaining the standard deviation spectrum (TVS)$^{1/2}$, the residual spectra were smoothed using
a wide Gauss filter $S$. The amplitude of the noise component decreases by a factor of $\approx\sqrt{\,S/h}$ after smoothing, where $h$ is the width of the pixel. If the width of the variable component is larger than the width of the filter, then smoothing will not significantly change the amplitude of the variable component, and peak in the standard deviation spectrum that corresponds to the variable component can be detected. 

This procedure, called smoothed Temporal Variance Spectrum (smTVS), was introduced by~\citet{Kholtygin2003}. 
The smTVS is described by next formula:
\begin{eqnarray}
\label{Eq.smTVS}
smTVS(\lambda,S)= \hspace{5cm} &    \nonumber  \\
=\frac{1}{N-1}
\left(  \sum\limits_{i=1}^{N} q_i(\lambda,S)\left[ F_i(\lambda, S) -
                     \overline{F(\lambda, S)}\,\right]^2  \right) \, , &
\end{eqnarray}
where $N$ is the number of spectra,  $F_{i}(\lambda,S)$ is the flux in the spectrum with number $i$ 
at wavelength $\lambda$ normalized to the continuum level and smoothed with variable Gauss filter 
($S$ is the filter width):
\begin{equation}
\centering
\label{Eq.smF}
F_i(\lambda, S) =   \frac{1}{\sqrt{2\pi} S}\int\limits_{-\infty}^{\infty} 
        F_i(l) e^{-\frac{1}{2}\left(\frac{l-\lambda}{S} \right)^2 }dl \, , 
\end{equation}
where $F_i(l)$ is the continuum normalized flux for spectrum $i$, 
$\overline{F(\lambda,S)}$ is the flux at wavelength~$\lambda$ averaged over all spectra:
\begin{equation}
\label{Eq.smFmean}
\overline{F(\lambda, S)} = \frac{1}{N}\sum\limits_{i=1}^{N} \omega_i F_i(\lambda, S)
                           {\left/ \sum\limits_{i=1}^{N}\omega_i \right.} \, .
\end{equation}
Here $\omega_{i}\sim (S/N)_i^2$ is the relative weight of {\it i}-th observation, where
  $(S/N)_i$ is the signal-to-noise ratio for spectrum $i$.

The wavelength correction factor $q_i(\lambda,S)$ which is accounted for pixel-to-pixel variations of the 
noise component within the line profile is considered in subsection 4.3. Outside the line $q_i(\lambda,S)\approx 1$, 
whereas within the line this factor can appreciably differ from $1$.

It was shown empirically that the best results are obtained for smoothing with a Gaussian filter width 
$S = 0.5-1.3\,$\AA{} (usually 15--30 pixels). For $S=0$ the value of smTVS($\lambda$,0) corresponds to 
the TVS value introduced by~\citet{Fullerton1996}.

Note that the efficiency of the method is sensitive to the number of the spectra, and it enhances substantially as this number increases.

For the sake of convenience smTVS can be normalized:
\begin{equation}
\label{Eq.smTVSnorm}
(smTVS)_{\mathrm{N}}(V,S)= \frac{smTVS(V,S)}{\left<smTVS_{\mathrm{cont}}(V,S)\right>}  \,.
\end{equation}
Here $\left<smTVS_{\mathrm{cont}}(V,S)\right>$ is the smTVS in the adjacent continuum near the line averaged 
over the interval $\Delta\lambda \gg h$, where $h$ is the pixel width. Similarly, we can introduce the normalized TVS: (TVS)$_\mathrm{N}$=(TVS)$_\mathrm{N}$(V)=(smTVS)$_{\mathrm{N}}(V,0)$.
 
\begin{figure}
  \includegraphics[width=1\columnwidth]{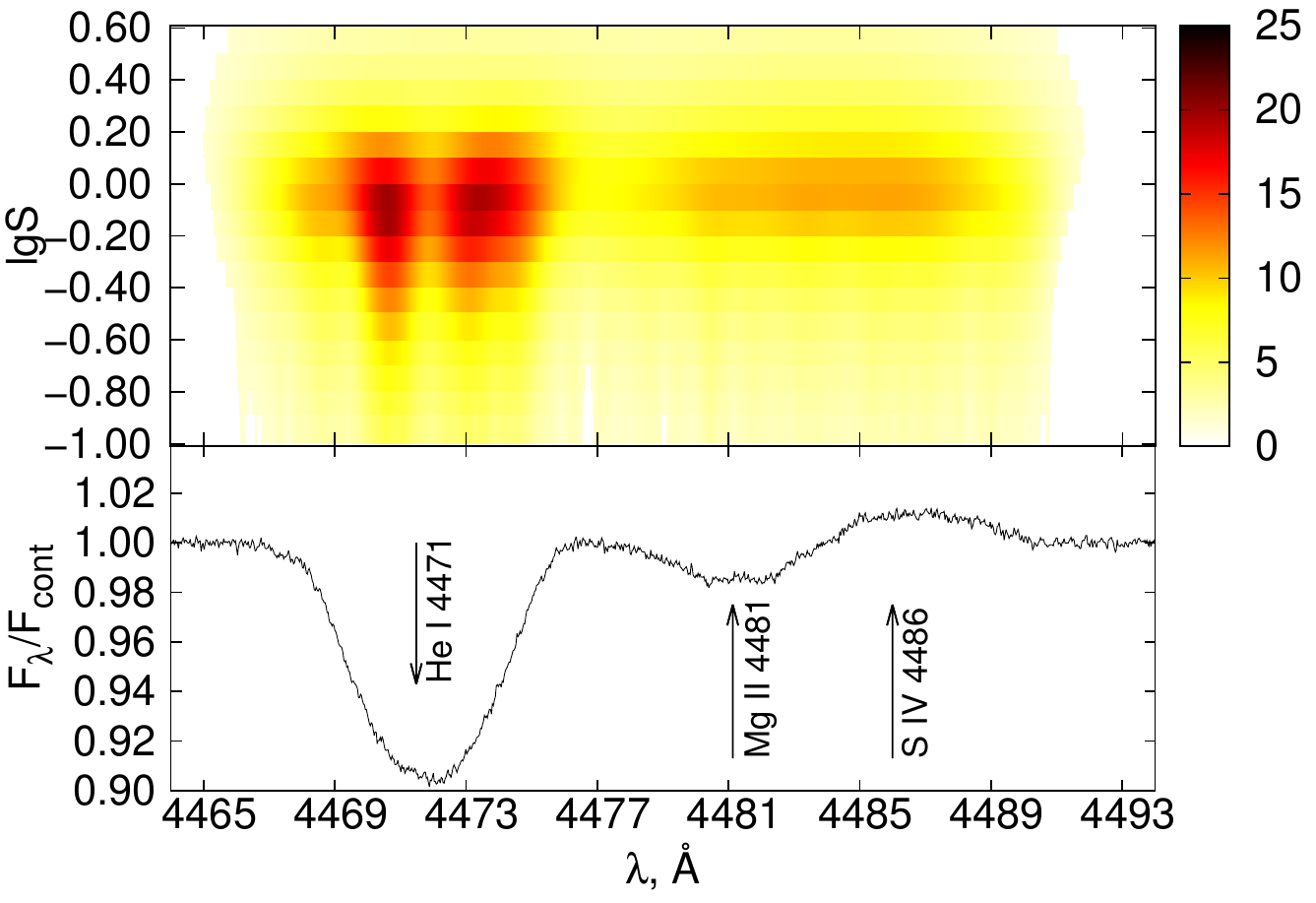}
  \caption{Density plot for (smTVS)$_{\mathrm{N}}^{1/2}$ (top) and mean 
           line profile of the line He\,I~$\lambda\,4471\,$\AA{} (bottom) in 
           the spectra of $\lambda\,$Cep.
           The darker areas correspond to higher amplitudes of the (smTVS)$_{\mathrm{N}}^{1/2}$. 
           Filer width S is expressed in decimal logarithms. 
           $S$ is measured in \AA.}
  \label{Fig.lCepmap}
\end{figure}

Fig.~\ref{Fig.lCepmap} presents a density plot of the (smTVS)$_{\mathrm{N}}^{1/2}$ for the 
He\,I~$\lambda\,4471\,$\AA{} line profiles in the spectra of $\lambda\,$Cep. 
Darker areas correspond to higher amplitudes of the smTVS. The density plot shows that the variability of 
the He\,I~$\lambda\,4471\,$\AA{} line is clearly presented at all filter widths. The smTVS also indicates 
variability of the weak Mg\,II~$\lambda\,4481\,$\AA{} and S\,IV~$\lambda\,4486\,$\AA{} 
lines profiles, which cannot be detected by TVS without smoothing.

\begin{figure}
  \includegraphics[width=1\columnwidth]{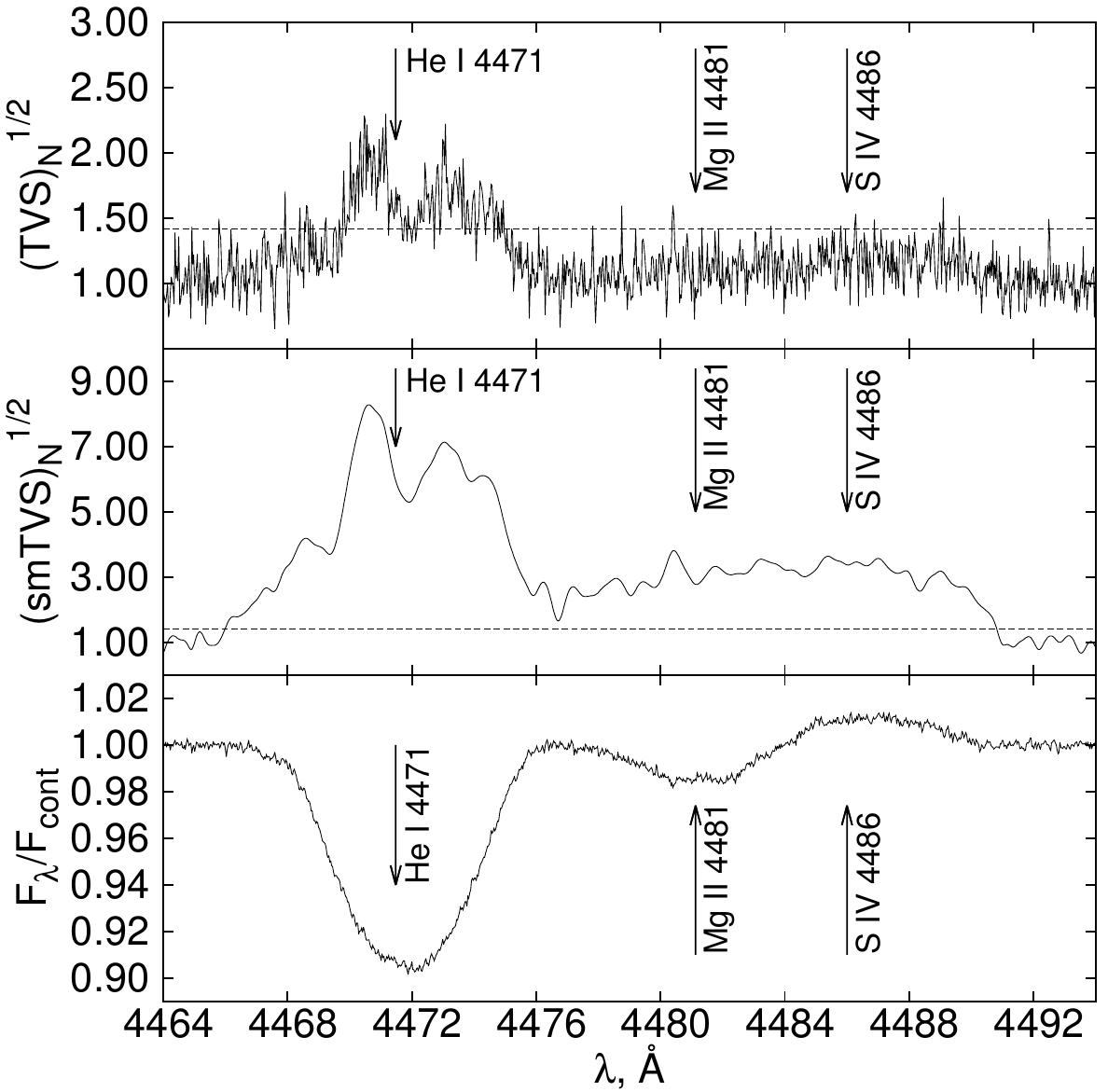}
  \caption{(TVS)$_{\mathrm{N}}^{1/2}$ (top),(smTVS)$_{\mathrm{N}}^{1/2}$ (middle)
           normalized by unity and mean line profile (bottom) 
           of the line He\,I~$\lambda\,4471\,$\AA{} in the spectra of $\lambda\,$Cep. 
           Filter width $S$ is 0.2\,\AA. 
           The horizontal line corresponds to the significance level 0.001.}
  \label{Fig.lCepTVS}
\end{figure}

In Fig.~\ref{Fig.lCepTVS} (TVS)$_{\mathrm{N}}^{1/2}$ (top panel) and (smTVS)$_{\mathrm{N}}^{1/2}$ (middle panel) functions for LPV in spectra of $\lambda\,$Cep are shown for comparison. The (smTVS)$_{\mathrm{N}}^{1/2}$ is a cross-section of the density plot in the Fig.~\ref{Fig.lCepmap} with the smoothing filter width 0.2\,\AA. The smTVS clearly demonstrates variability of the Mg\,II and S\,IV lines which is undetected by simple TVS analysis. The variability of the He\,I line is seen in TVS and smTVS both. The smTVS shows that whole He\,I line profile is variable though the TVS detects the variability in central part of the line profile.

Note that in spite of the fact that a lot of lines are extremely weak and their residual intensities do not exceed the noise level of the adjacent continuum their variability is clearly detected in the case of using smTVS. 

Unfortunately, the smTVS technique cannot be used for accurate localization of weak lines with variable profiles for large filter widths.

\section{Statistical properties of the smTVS}

Let $\Xi = (\xi_1, \xi_2, \dots , \xi_N$) be the set of $N$ mutually independent normally distributed random values  
with means $\mu_1, \mu_2, \dots ,\mu_N$ and variances $\sigma^2_1, \sigma^2_2, \dots , \sigma^2_N$. 
For $N$ constants $a_1,  a_2, \dots , a_N$  we can introduce the random variable
\begin{equation}
\label{Eq.LinComnRND}
\zeta= \sum\limits_{i=1}^N a_i \xi_i \,.
\end{equation}
Then accordingly for example~\cite{Brandt1970} the variable $\zeta$ has a normal distribution with mean and variance 
\begin{equation}
\label{Eq.LinComnRND_Mean}
\mathbf{E}[\zeta] = \sum\limits_{i=1}^N a_i \mu_i,  \quad \mathbf{Var}[\zeta] = \sum\limits_{i=1}^N a_i^2 \sigma^2_i \, .
\end{equation}

In a simple case $\mu_1 = \mu_2 = \dots = \mu_N = \mu$  and $\sigma^2_1 = \sigma^2_2 = \dots = \sigma^2_N = \sigma^2$
the mean and the variance
\begin{equation}
\label{Eq.LinComnRNDsym_MeanVar}
\mathbf{E}[\zeta] = \mu \sum\limits_{i=1}^N a_i, \quad \mathbf{Var}[\zeta] = \sigma^2  \sum\limits_{i=1}^N a_i^2 \, .
\end{equation}
Filtering of the signal can completely or partially suppress some aspect of the signal \cite[e.g.,][]{Orfanidis2010}.
The operator of filtering $\hat{S}$ is the convolution of the source (signal) function $f(x)$ with the family of 
the filter functions $G_S(x)$ having a variable width~$S$:
\begin{equation}
\label{Eq.Filtering}
\hat{S}[f](z) = \int\limits_{-\infty}^{-\infty} G_S(z-x) f(x) dz. 
\end{equation}
Here the filter function  $G_S(x) = (1/s)g(x/S)$.
Hereinafter we will use both the simple {\large\it rectangular} filter: 
$g_{\mathrm{R}}(x)= 1$ for $x\in [-1/2,1/2]$ and $g_{\mathrm{R}}(x)= 1$ for all other values of $x$,
and the {\large\it Gauss} filter $g_{\mathrm{N}}(x) =  (2\pi)^{-1/2}\exp({-x^2/2})$. 

For smoothing the function $f(x)$ we have to calculate the integral Eq.~(\ref{Eq.Filtering}).
Suppose that $f(x)$ is determined in the points $\{ x_i \}$, where $i= 0,1,\dots , N$. Then we can present the 
integral Eq.~(\ref{Eq.Filtering}) as a sum
\begin{equation}
\label{Eq.NumerFilt}
\hat{S}[f](z) \approx \sum\limits_{i=1}^{N-1} a_i f(\tilde x_i)\,.
\end{equation}
Here $a_i = G_S(z - \tilde x_i)$, where $\tilde x_i = x_i+ \varepsilon\, \Delta x_i$, $\Delta x_i = x_{i+1}-x_i$ 
and the parameter $\varepsilon \le 1$.

\subsection{Smoothing the random function}

Suppose that the random function $f(x)$ is determined at the points $x_i, i=1,2,\dots,N$  and for each $i$ 
the value of $f(x_i)$ is the normally distributed random value $\xi_i$ with the mean $\mu$, the variance $\sigma^2$, 
and  with the distribution function ${\cal F}(x_i)$. Smoothing the random function $f(x)$ with using the arbitrary 
operator of filtering $\hat{S}$ gives us the next expression:
\begin{equation}
\label{Eq.SmRandom}
\eta = \eta(z) =\hat{S}[\zeta] \approx \sum G_S(z-\tilde x_i) \Delta x_i \xi_i \, .
 \end{equation}
If the distribution function ${\cal F}(x_i)$ does not depend on $i$ then the sum (\ref{Eq.SmRandom}) does not 
depend on the value of $z$. For the sake of the convenience we can put $z=0$. The random value $\eta$ is 
the normally distributed with the mean and the variance which can be found accordingly Eq.~(\ref{Eq.LinComnRND_Mean}).

\begin{figure}
\centering
  \includegraphics[width=0.8\columnwidth]{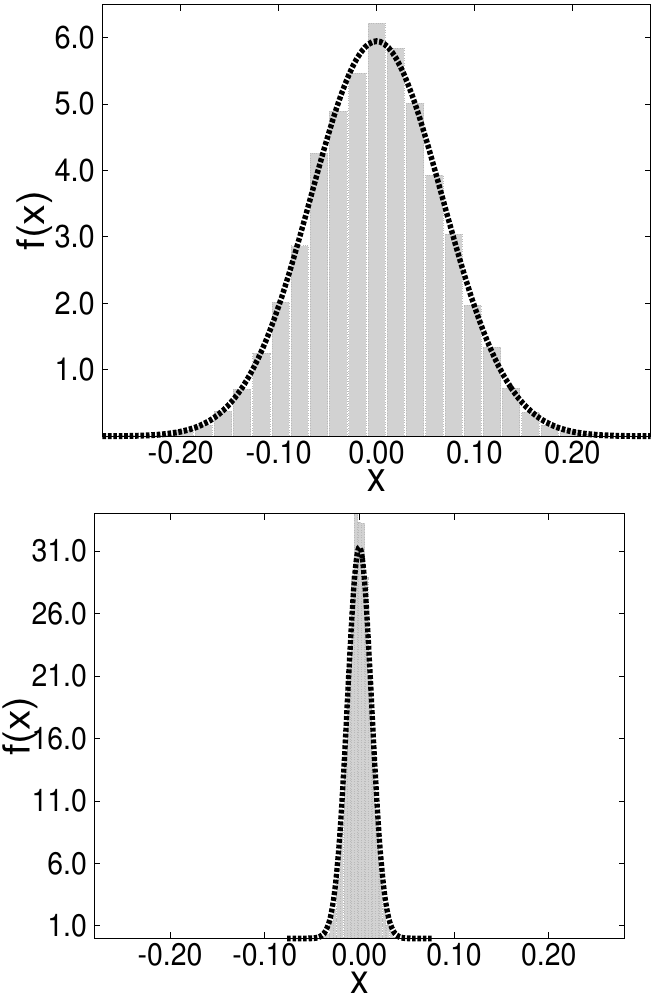}
  \caption{{\it Top panel}: distribution function for the smoothed with the rectangular filter 
          normally distributed random variables with $\mu=0$, $\sigma=0.6$ and $h=0.01$ for 
          the filter width $S=0.8$. The thick dashed line shows the Gauss fit.
           {\it Bottom panel}: the same as in the top panel, but for the Gauss Filter.}
  \label{Fig.RandSmRectGauss}
\end{figure}

Hereinafter for simplicity we will suppose that $\varepsilon = 1/2$, then  $\tilde x_i = (i+1/2)h$, 
where $i=0,\pm 1, \pm 2, \dots$. In this case for all $i$ $\Delta x_i= h$.
Then
\begin{equation}
\label{Eq.SmRandom-h}
 \eta = \hat{S}[\zeta] =  \sum\limits_{i=-\infty}^{\infty} G_S(ih+ h/2) h \xi_i =  
        \sum\limits_{i=-\infty}^{\infty} a_i \xi_i \, ,
 \end{equation}  
where $a_i = G_S(ih+ h/2) h$. Suppose that for all $i$ the mean $\mu_i = \mu$  and the variance $\sigma^2_i = \sigma^2$. Then the mean and variance of $\eta$
\begin{equation}
\label{Eq.MeanVar_SmRandomVal}
\mathbf{E}[\eta] \approx \mu \sum\limits_{i=-\infty}^{\infty} a_i \mu_i   = \mu , \quad
\mathbf{Var}[\eta] \approx \sigma^2 \sum\limits_{i=-\infty}^{\infty} a_i^2  \, .
\end{equation}
For the symmetrical filters $\mathbf{Var}[\eta] = \sigma^2 \sum_{i=0}^{\infty} 2 a_i^2$.
In the case of the rectangular filter it is easy to find that
\begin{equation}
\label{Eq.MeanVar_SmRect}
\mathbf{E}[\eta] = \mu , \:\: \mathbf{Var}[\zeta] = \sigma^2\left(\frac{h}{S}\right) , \:\:
\mathbf{St.Dev}[\eta] = \sigma  \sqrt{\frac{h}{S}}  .
\end{equation}
In Fig.~\ref{Fig.RandSmRectGauss} (top panel) we plot the distribution of the smoothed with the the rectangular filter ($S=0.8$) the normally distributed values $\xi_i$ with parameters $\mu=0$, $\sigma=0.5$ and  
$h=0.01$. We can see that in accordance with Eq.~(\ref{Eq.MeanVar_SmRect}) the smoothed variable is also normally 
distributed but with the standard deviation which is $\approx \sqrt{0.8/0.01}\approx 9$ time smaller. 

For the the scaled Gauss filter with width $S$ the mean of the random value 
$\eta$ is $\mathbf{E}[\eta] = \mu$. 
To calculate the variance we can use Eq.~(\ref{Eq.MeanVar_SmRandomVal}), where
\begin{equation}
\label{Eq.a_i_SmGauss}
a_i =  \frac{h}{S\sqrt{2\pi}}e^{-\frac{1}{2}\left(\frac{ih+h/2}{S}\right)^2} \, ,
\end{equation}
Substituting Eq.~(\ref{Eq.a_i_SmGauss}) in Eq.~(\ref{Eq.MeanVar_SmRandomVal}) one can find that
\begin{equation}
\label{Eq.Var_SmGauss}
\mathbf{Var}[\eta] =  
        \left(\frac{\sigma^2 }{\pi}\right) \left(\frac{h}{S}\right)^2 \times \sum\limits_{i=0}^{\infty} 
        e^{-\left(\frac{h}{S}\right)^2 (i+1/2)^2}
\, .
\end{equation}
Denote $r=(h/S)$, $q=\exp{(-r^2)}$, then
\begin{equation}
\label{Eq.Var_Gauss_bi}
\mathbf{Var}[\eta] = \frac{\sigma^2 r^2}{2\pi} \times \sum\limits_{i=0}^{\infty} 2 q^{(i+1/2)^2} \, .
\end{equation}
Here $\sum_{i=0}^{\infty} 2q^{(i+1/2)^2} = \theta_2(0,q)$ (e.g. \cite{Yanke-1960}). Thus
\begin{equation}
\label{Eq.Var_SmGauss_theta2}
\mathbf{Var}[\eta] = \frac{\sigma^2 r^2}{2\pi} \times \theta_2(0,q) \, . 
\end{equation}
Usually the value of $r^2= (h/S)^2 \ll 1$. Thus
$q\approx 1$. In this case we can use the asymptotic by \cite{Zhuravskii1941} of the function $\theta_2$ at $q\to 1$:
$\theta_2(0,q) \approx \sqrt{\rho'} \left(1 - 2q' + \dots \right)$, where 
$\rho'= \pi/ \ln({1/q})=\pi/r^2$ and $q'=\exp(-\pi \rho')=\exp(-(\pi/r)^2)$. 
Using this asymptotic one can easily find that at $q\to 1$
\begin{equation}
\label{Eq.Var_Gauss_fin}
\mathbf{Var}[\eta] \approx \frac{1}{2\sqrt{\pi}}\sigma^2 \left(\frac{h}{S}\right)\left(1 - 2e^{-(\pi/r)^2} + \dots \right) \, ,
\end{equation}
and
\begin{equation}
\label{Eq.StDev_Gauss_fin}
\mathbf{St.\,Dev.\,}[\eta] \approx \frac{\sigma}{2^{1/2}\pi^{1/4}}\, \sqrt{\frac{h}{S}}\left(1 - e^{-(\pi/r)^2} + \dots \right) \, .
\end{equation}
%
\begin{figure}
\centering
   \includegraphics[width=0.8\columnwidth]{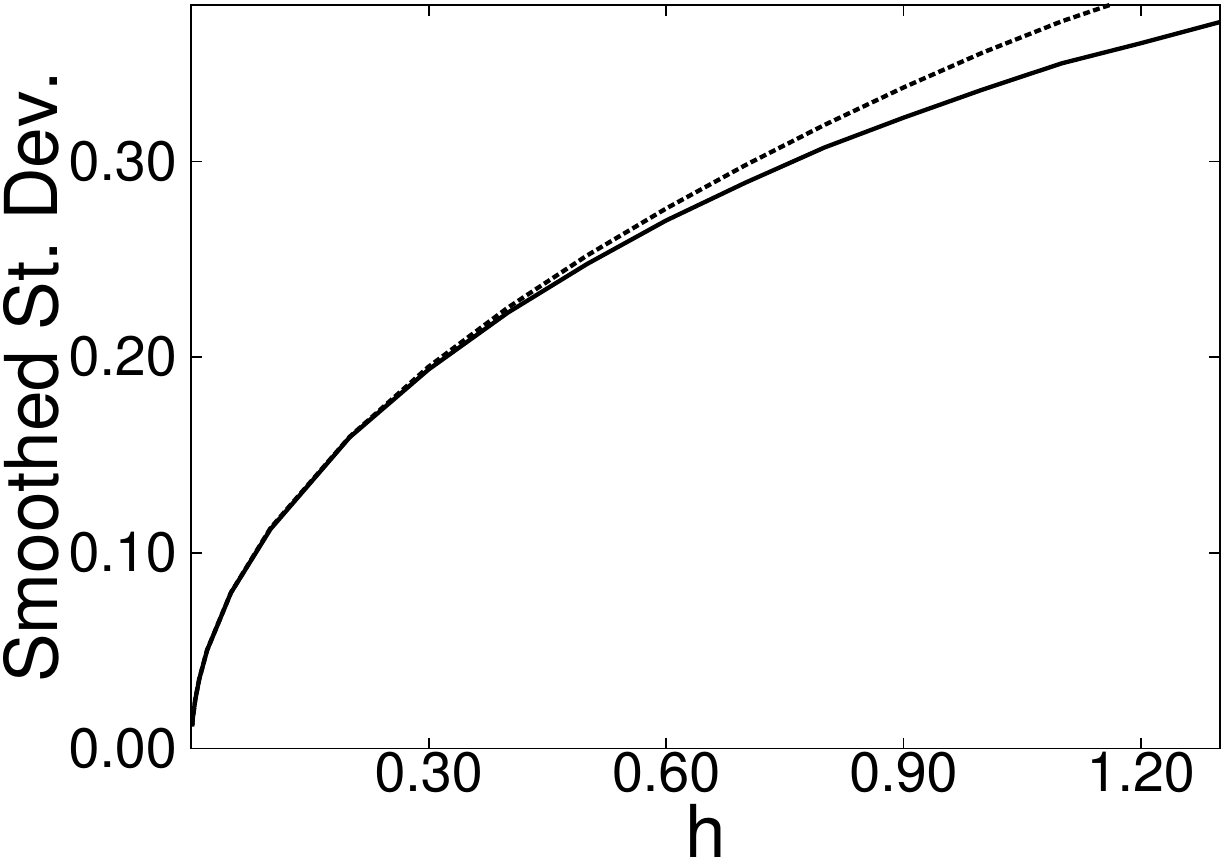}
   \caption{Standard Deviation for the smoothed with the Gauss filter ($S=0.8$) normally distributed random variables 
            with $\mu=0$, $\sigma=0.6$ as a function of $h$ (solid line). The analytical fit 
            Eq.~(\ref{Eq.StDev_Gauss_fin}) is shown with the dashed line.}
\label{Fig.StDev_GaussCalcAnal}
\end{figure}

Thus, in the case of the Gauss filter the variance of the smoothed random variables for the same filter width 
is much smaller than in the case of the rectangular one. This conclusion is illustrated in Fig.~\ref{Fig.RandSmRectGauss} (bottom panel) where the distribution of $\eta$ smoothed with the Gauss filter ($S=0.8$) is the normally distributed values $\xi_i$ with parameters $\mu=0$, $\sigma=0.6$ for $h=0.01$ is given. 

The quality of the fit Eq.~(\ref{Eq.StDev_Gauss_fin}) for all ratios  $h/S<0.5$ is very good as it clearly shown
in Fig~\ref{Fig.StDev_GaussCalcAnal}. It means that in all practically important cases we can use 
Eq.~(\ref{Eq.StDev_Gauss_fin}) to obtain the Standard Deviation for the smoothed with Gauss filter random values.

\subsection{Correlation of the smoothed random variables}

\begin{figure}
\centering
   \includegraphics[width=0.8\columnwidth]{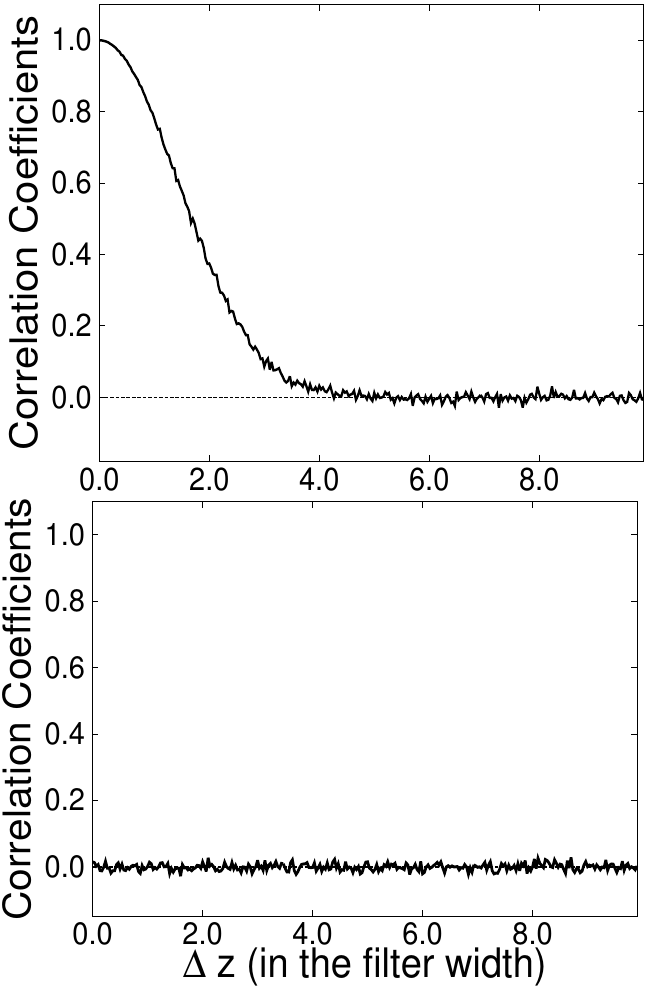}
   \caption{{\it Top panel:} the correlation coefficients for the smoothed with the Gauss filter ($S=0.2$) 
            normally distributed random variables as a function of $\Delta z$ (solid line) for $\mu=0$, $\sigma=0.5$ 
            and $h=0.05$ for one set of random values. 
            The level $r=0$ is marked with the dashed line.
            {\it Bottom panel:} the same as in the top panel, but for random variables from different sets.}
\label{Fig.CorcCoeff_smRandoms}
\end{figure}

Smoothed with an arbitrary filter, random value $\eta(z)$ (see Eq.~(\ref{Eq.SmRandom})) depends on the value 
of $z$. Thus, for small difference $\Delta z = |z_1 - z_2|$ the random values $\eta(z_1)$ and $\eta(z_2)$ are not
independent. The correlation coefficients $r=r(z_1,z_2)$ depends on the difference $\Delta z$ and the width of the filter $S$.

In Fig.~\ref{Fig.CorcCoeff_smRandoms} (top panel) we plot the value of $r$ as a function of $\Delta z$ (in units 
of the filter width). Inspecting the figure we can conclude that only for differences $\Delta z >(3-4)\times S$ 
the random variables $\eta(z_1)$ and $\eta(z_2)$ become independent.

On the other hand if we consider  different sets $\Xi$ and $\Theta$ of random values with 
equal means and variances, then the smoothed random variables referred to these sets are independent as clearly shown in Fig.~\ref{Fig.CorcCoeff_smRandoms} (bottom panel).

\subsection{Statistics of the smTVS}

Let $F_i(\lambda)$ be the continuum normalized flux at the wavelength $\lambda$ for $i$-th spectrum. It can be presented as sum
\begin{equation}
\label{Eq.Flux-plus-noise}
F_i(\lambda) = I_{\mathrm{c}}+ Z_i(\lambda)+ N_i(\lambda) \, ,  
\end{equation}
where $I_{\mathrm{c}}=1$, $Z_i(\lambda)$ is the net continuum normalized line flux at the wavelength $\lambda$ 
and $N_i(\lambda)$ is the corresponding noise contribution. Out of line profile $Z_i(\lambda)=0$.

Suppose that the noise component $N_i(\lambda)$ can be described by the Gauss (normal) 
distribution $N(0,\sigma_{i}(\lambda))$, where $\sigma_{i}(\lambda)$ is the standard deviation 
for the noise component for $i$-th spectrum at wavelength~$\lambda$.  Outside the line 
$\sigma_{i}(\lambda)= \sigma_{ic}$, where $\sigma_{ic}$ is the noise standard deviation in the continuum near 
the line. For weak lines we can put that the noise $\sigma_{i}(\lambda)=\sigma_{ic}$ both inside and outside the line.

Analogously Eq.~(10) by \cite{Fullerton1996} we can introduce the mean noise value $\sigma_0$ for considered $N$ spectra:
\begin{equation}
\label{Eq.sigma0}
\sigma_0 =  \left(\frac{1}{N}\sum\limits_{i=1}^{N} \left(\sigma_{ic}\right)^{-2} \right)^{\!-1/2} \, .
\end{equation}

The ratio $\sigma_{i}(\lambda)/\sigma_{ic}$ can be calculated using equation~(8) by \cite{Fullerton1996}.
Taking into account Eq.~(\ref{Eq.Flux-plus-noise}) we easily can obtain that 
\begin{equation}
\label{Eq.sigma0}
\alpha_{i}(\lambda) = \left(\!\frac{\sigma_{i}(\lambda)}{\sigma_{ic}}\!\right)^2 = 
         \frac{1 + Z_i(\lambda) + (1+ 1/b)g'R^2}{1 + (1+1/b)g'R^2}  \, .
\end{equation}
Here $b$ is the number of bias exposures averaged, $R$ is the read-out noise in analog-to-digital conversion unit (ADU) and $g'=g/E_{ic}$, where $g$ is the number of electrons per ADU and $E_{ic}$ is the continuum flux in ADU.

The factor $\alpha_{i}(\lambda)$ is responsible for pixel-to-pixel variations of $\sigma_{i}(\lambda)$ 
arising due to the presence of absorption ($Z_i(\lambda)<0$) or emission ($Z_i(\lambda)>0$) features. 
In the limit of relatively high signal-to-noise ratio $S/N>50$ typical for studied LPV and for usual value of $R=$2--3 the read-out noise can be neglected and $\alpha_{i}(\lambda)=1+Z_i(\lambda)$.

Resuming we can write the wavelength correction factor in Eq.~(\ref{Eq.smTVS}) for $S=0$ as
\begin{equation}
\label{Eq.q_i(lambda)}
q_i(\lambda,0) = \frac{\omega^{'}_i}{\alpha_{i}(\lambda)} = \frac{\sigma^2_0}{\sigma^2_i(\lambda)} \, ,
\end{equation}
where $\omega^{'}_i=\omega_i/\sum\limits_{i=1}^N\omega_i$. Taking into account that $\sigma_{ic}=(S/N)_i^{-1}$, where $(S/N)_i$ is the signal-to-noise ratio for spectrum $i$ and using 
equation~(\ref{Eq.sigma0}) one can find that $\omega^{'}_i = (\sigma_0/\sigma_{ic})^2$ (cf. Eq.(11)-(13) by \cite{Fullerton1996}).

The smoothed flux can be described as
\begin{equation}
\label{Eq.Flux-plus-noiseSm}
F_i(\lambda,S) = I_{\mathrm{c}}+ Z_i(\lambda,S)+ N_i(\lambda,S) \, ,   
\end{equation}
where  the smoothed net line profile $Z_i(\lambda,S)$ can be evaluated using Eq.~(\ref{Eq.smF}) with replacing 
$F_i(l)$ by $Z_i(l)$. The smoothed noise component 
\begin{equation}
\label{Eq.Flux-plus-noiseSm}
N_i(\lambda,S) = \frac{1}{\sqrt{2\pi} S}\int\limits_{-\infty}^{\infty} 
        N_i(l) e^{-\frac{1}{2}\left(\frac{l-\lambda}{S} \right)^2 }dl\,,
\end{equation}
where $N_i(l)$ has the normal distribution $N(0, \sigma_i(l))$ with $\sigma_i(l)=\sigma_{ic}\sqrt{1+Z_i(l)}$\,.
Here we understand the integration of the random value of noise $N_i(l)$ in terms of calculating the integral sum of random variables like it was considered in Eq.~(\ref{Eq.SmRandom}).

Outside the line the smoothed noise component $N_i(\lambda,S)$ has the $N(0, \sigma^{\mathrm{sm}}_{ic})$ 
distribution, where the value of $\sigma^{\mathrm{sm}}_{ic}$ can be 
obtained from  Eq.~(\ref{Eq.StDev_Gauss_fin}) with the replacement of $\sigma$ by $\sigma_{ic}$ and taking into account 
that parameter $h$ is the pixel width. Thus 
\begin{equation}
\label{Eq.sigmaic^sm}
\sigma^{\mathrm{sm}}_{ic} \approx 
\frac{\sigma_{ic}}{2^{1/2}\pi^{1/4}}\, \sqrt{\frac{h}{S}}\left(1 - e^{-(\pi S/h)^2}\right) \, .
\end{equation}
Due to the strong dependence $\sigma_i(\lambda)$ on $\lambda$ the distribution function for the smoothed 
noise $N_i(\lambda,S)$ within and near the line cannot be found analytically.           

Let us describe the smoothed residual spectra by the usual way (see Eq.~(\ref{Eq.smTVS})):
\begin{equation}
\label{Eq.DiffSpectra}
d_i(\lambda,S) = F_i(\lambda,S) - \overline{F_i(\lambda,S)}\, .
\end{equation}
The value $d_i(\lambda,S)$ corresponds to the residual spectra $d_{\ij}$ by \citet{Fullerton1996} 
(see their Eq.~2) but for smoothed fluxes $F_i(\lambda,S)$. 
Consider the random variable
\begin{equation}
\label{Eq.zeta_sm}
\eta = \eta(\lambda,S) = \frac{1}{N-1}\sum\limits_{i=1}^{N} q_i(\lambda,S) d^2_i(\lambda,S) = 
       \frac{1}{N-1}\sum\limits_{i=1}^{N} \xi^2_i \, ,
\end{equation}
where $q_i(\lambda,S) = \left(\sigma^{\mathrm{sm}}_0/\sigma_{i}(\lambda,S)\right)^2$. The parameter  
\begin{equation}
\label{Eq.sigma0sm}
\sigma^{\mathrm{sm}}_0 =  \left(\frac{1}{N}\sum\limits_{i=1}^{N} \left(\sigma^{\mathrm{sm}}_{ic}\right)^{-2} \right)^{-1/2}
\end{equation}
is an analogue of value $\sigma_0$ but for smoothed fluxes. The random value 
$\xi_i=\left(q_i(\lambda,S)\right)^{1/2} d_i(\lambda,S)$. For all $i$ $\mathbf{E}[\xi_i]=0$. Thus $\mathbf{E}[\eta]=0$. Any random value $\xi_i$ has $N(0,\sigma^{\mathrm{sm}}_0)$ distribution. 

\begin{figure}
\centering
  \includegraphics[width=0.9\columnwidth]{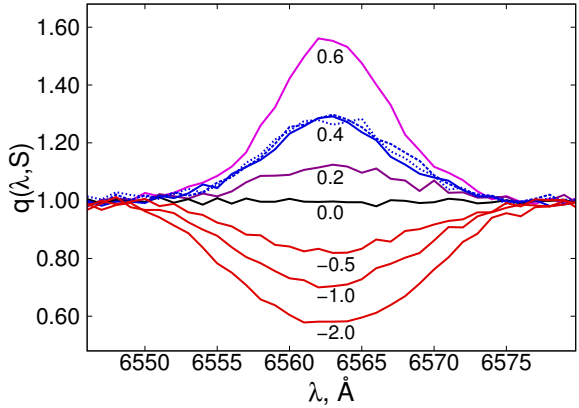}
  \caption{The factor $q(\lambda,S)$ for line H$_{\alpha}$ and for the different values of the parameter $A$ 
           (marked at the corresponding curves) for absorption ($A > 0$, blue and purple curves) 
           and emission ($A < 0$, red curves) line profiles for $S=0.8\,$\AA{} and $S/N=50$. 
           Blue solid, dashed and dotted lines show the variations of the parameter $q(\lambda,S)$ for $A=0.4$ and 
           $S=0.0,\,0.8,\,\mathrm{and}\,2.4\,$\AA{} respectively.
           }
  \label{Fig.qLSlineArr}
\end{figure}

To illustrate, we present the results of our calculations of the 
selected curves $q(\lambda,S)$ for the model H$_{\alpha}$ net line profile in Fig.~(\ref{Fig.qLSlineArr}):
\begin{equation}
\label{Eq.Zmodel}
Z(\lambda)= -A e^{-\frac{1}{2} \left(\frac{\lambda-\lambda_0}{W}\right)^2 } \, ,
\end{equation}
and the signal-to-noise ratio $S/N=50$. Here $\lambda_0=6562.849\,$\AA\ is the laboratory wavelength of the line and 
$W=5\,$\AA\ is the typical value of the H$_{\alpha}$ line width in the spectra of early type stars correspondingly. 
The value of $A$ is the line depth (positive for the absorption line and negative for the 
emission ones). Inspecting Fig.~(\ref{Fig.qLSlineArr}) it is possible to conclude that the factor $q(\lambda,S)$ 
has to be taken into account mainly for the strong ($|A|>0.3$) lines. 

Note that the variation of the parameter $S$ and the ratio $S/N$ extremely weakly influence the value of $q(\lambda,S)$. This result holds for different values of the line profile parameters and the ratio $S/N$.
Therefore for all $S$ we can put $q(\lambda,S)\approx q(\lambda,0)$. Thus for all values of $S$ 
it is possible to use the simple formula~(\ref{Eq.q_i(lambda)}) instead of making the numerical integration in  Eq.~(\ref{Eq.Flux-plus-noiseSm}).

The smoothed random values $\xi_i$ are independent as belonging to different spectra (see subsection 4.2).
Thus the random value $\eta(\lambda,S)$ (smooth Temporal Variance Spectrum) has 
$(\sigma^{\mathrm{sm}}_0)^2 \chi^2_{N-1}$ distribution. 

Next we determine by the usual way the small value of the significance level $\alpha\ll 1$ and a value 
$\chi^{2}_{\alpha}$ such as the probability  $P(\chi^2 >\chi^{2}_{\alpha}) = \alpha$ for given $N-1$ degrees 
of freedom. We suppose if within the line the value 
$smTVS(\lambda,S) > \left(\sigma^{sm}_0\right)^2 \chi^{2}_{\alpha}/(N-1)$ 
then the LPV is real at the significance level $\alpha$. 

Our calculations show that with the error smaller than $10^{-3}$ in the continuum units the value 
$\left<smTVS_{\mathrm{cont}}(\lambda,S)\right> = \left(\sigma^{sm}_0\right)^2$.  It  signifies that normalized 
$(smTVS)_{\mathrm{N}}(\lambda,S)= smTVS(\lambda,S)/\left(\sigma^{sm}_0\right)^2$. In the case of using 
the normalized smTVS we will suppose that LPV is real if $smTVS_{\mathbf{N}}(\lambda,S) > \chi^{2}_{\alpha}/(N-1)$.

\section{Calculations of line profiles in spectra of non-radial pulsating stars} 

The line profile of non-radial pulsating star in an arbitrary moment of time $t$ can be represented as
\begin{equation} 
\label{Eq.I_NRP}
I(\lambda) =  I_{\mathrm{mean}}(\lambda,t) + \Delta I (\lambda,t) \, . 
\end{equation} 
Here $I_{\mathrm{mean}}(\lambda)$ is the main part of the line profile that does not depend on time (average profile), $\Delta I(\lambda)$ is the variable part of the profile that can be presented according to~\citet{Telting1997III} as follow:
\begin{eqnarray} 
\label{Eq.DeltaI_NRP}
\Delta I(\lambda)   = & I_{0}(\lambda)\sin(\omega t + \Psi_0(\lambda)) &   \nonumber      \\ 
                      &  +  I_{1}(\lambda)\sin(2\omega t + \Psi_1(\lambda))&              \\
                      &  +  I_{2}(\lambda)\sin(3\omega t + \Psi_2(\lambda))&   + \dots\,. \nonumber  
\end{eqnarray} 

\citet{Telting1997III} performed detailed modelling of line profiles in the spectra of non-radially
pulsating stars using Monte-Carlo method for $l\le 15$, $|m|\le l$ and random values of parameters 
$i,W,V_{\mathrm{max}},k, \omega,\Omega$ etc. Here $i$ is the inclination of the rotational axis, $W$ is the width 
of the intrinsic line profile for any surface element of the star, $V_{\mathrm{max}}$ is the maximal pulsation 
velocity amplitude, $k$ is defined as the ratio of the horizontal to the vertical pulsation velocity amplitude, $\omega$ is the angular pulsation frequency and $\Omega$ is the angular rotation frequency of the star. Their calculations showed that amplitudes of harmonics $I_i(V)$ decrease fast along with increasing number of harmonic and the phase difference between red and blue wings is  $\Delta\Phi_i=\Phi_i(+V\sin i) - \Phi_i(-V\sin i) \sim \pi (i+1) l$. The authors found out that the value  $\Delta\Phi_0$ is determined by $l$, and the value $\Delta\Phi_1$ is determined by modulus of the value $m$.

\citet{Telting1997III} obtained the relations:
\begin{eqnarray} 
\label{Eq.lm-DeltaPsio1_NRP}
                   l   \approx  & p_0 + q_0 \frac{|\Delta\Phi_0|}{\pi} = ~0.10 + 1.09 \frac{|\Delta\Phi_0|}{\pi}  \, , &           \\ 
                  |m|  \approx  & p_1 + q_1 \frac{|\Delta\Phi_0|}{\pi} = -1.33 + 0.54 \frac{|\Delta\Phi_1|}{\pi}  \, , &       
\end{eqnarray} 
that can be used for modelling line profiles in the spectra of non-radial pulsating stars.

The modelling shows that the component $I_{2}$ practically does not affect the profile shape. Therefore only main frequency ($I_{0}$) and its first harmonic ($I_{1}$) can be used in equation~(\ref{Eq.DeltaI_NRP}).

The dependences $I_0(V)$ and $I_1(V)$ obtained 
in~\citet{Schrijvers1997, Telting1997III} we have approximated as 
\begin{eqnarray} 
\label{Eq.I_0(V)-I_1(V)}
     I_0(V)   = & I^0 (1-x^{\mu_1}), &   \nonumber         \\ 
     I_1(V)   = & I^1 (1-x^{\mu_2}), &  
\end{eqnarray} 
where $I^0=I_0(0)$,  $I^1=I_1(0)$, the parameter $x=\left|V/(V\sin i+W)\right|$.
The values of $\mu_1$ and $\mu_2$ are chosen so that they fit the dependences $I_0(V)$ and $I_1(V)$ 
by \citet{Schrijvers1997}. Our computations show that the values of $\mu_1 \approx \mu_2 =$~3--4 are optimal. 

The amplitudes of harmonics of main pulsation frequency $\nu = 2\pi\omega$ decrease fast with increasing the number of harmonic $j$. Usually $I^j/I^{j+1} =$~3--5.  

The phase dependences of the main component and its first harmonic consistent with empirical dependences~(\ref{Eq.lm-DeltaPsio1_NRP}) can be written as
\begin{eqnarray} 
\label{Eq.Psi_0(V)-Psi_1(V)}
     \Psi_0(V)   = & \Psi_0(V) + \gamma \frac{l  -p_0}{q_0} \frac{\pi}{2}, &   \nonumber         \\ 
     \Psi_1(V)   = & \Psi_0(V) + \gamma \frac{|m|-p_1}{q_1} \frac{\pi}{2} . &  
\end{eqnarray} 
The parameter $\gamma$ defines the direction of motion of pulsation waves on the stellar surface.The value $\gamma=-1$ corresponds to prograde (p) waves that move in the direction of star rotation. The value $\gamma=1$ corresponds to retrograde (r) waves that move in the direction opposite to the rotation of the star.

\begin{figure}
  \includegraphics[width=1\columnwidth]{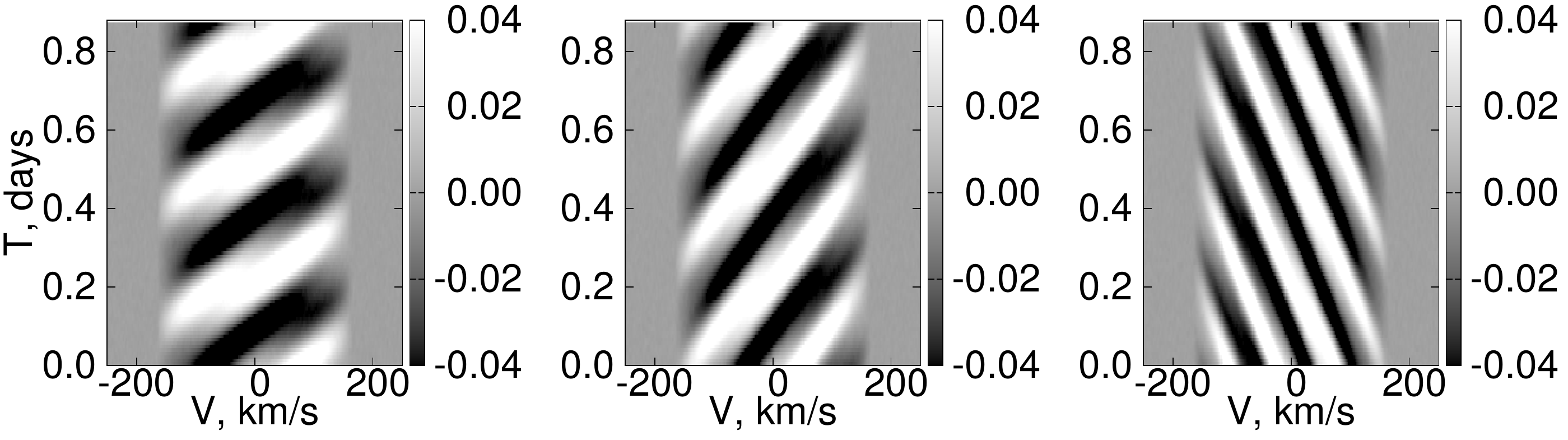}
  \caption{{\it Left panel}: model residual profiles for pulsation mode $l=2$, $m=-2$, $\gamma=-1$ 
           for three pulsation cycles. 
           {\it Middle panel}: the same as in the left panel but for $l=4$, $m=-4$. 
           {\it Right panel}: the same as in the left panel but for $l=8$, $m=8$, $\gamma=1$.}
  \label{Fig.NRPmap}
\end{figure}

The results of the modelling of residual profiles in sectoral $(l=m)$ mode using equations~(\ref{Eq.DeltaI_NRP}), (\ref{Eq.I_0(V)-I_1(V)}) and~(\ref{Eq.Psi_0(V)-Psi_1(V)}) are presented in Fig.~\ref{Fig.NRPmap}. Fig.~\ref{Fig.NRPmap} clearly shows the motion of the features along the line profile from blue to red wings in a case of prograde motion and from red to blue for retrograde motion and that with increasing $l$ the number of the features, related to NRP, increases. With increasing $l$ the velocity of motion of the NRP features along the line profile also increases.

\begin{figure}
\centering
  \includegraphics[width=0.8\columnwidth]{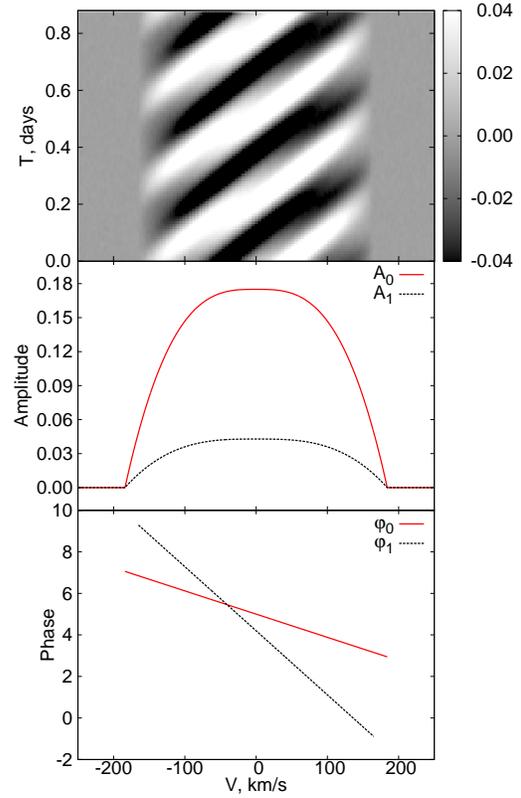}
  \caption{Line profile variations due NRP in mode (4,-4) for the set of parameters 
           similar to those used by \citet{Telting1997II} in their Fig.~1. 
           From top to bottom: residual profiles 
           for 3 pulsation cycles, distribution of the amplitudes of the variations 
           with input pulsation frequency $I_{0}$ (solid red line) and its first 
           harmonic $I_{1}$ (dashed black line) expressed in units of average central 
           line depth according to~\citet{Telting1997II}, phase distribution of 
           the variations with input pulsation frequency $\Phi_0$ (solid red line) 
           and its first harmonic $\Phi_1$ (dashed black line) expressed in units
           of $\pi$ radians.}
  \label{Fig.TS}
\end{figure}

The calculations for NRP mode (4,-4) with the set of parameters as used by \citet{Telting1997II} in their 
Fig.1 were done to check our method of modelling NRP. From a comparison of our Fig.~\ref{Fig.TS} and Fig.1 
by \citet{Telting1997II} we can conclude that both amplitude and phase curves in both cases are 
close, but the differences are noticeable. These differences have a little effect on the LPV, because 
the shape of the mean profile has been removed from the individual ones.
%

\section{Results and discussion} 
\subsection{Diagnostic of ultra weak LPV}

\begin{figure}
\centering
  \includegraphics[width=0.9\columnwidth]{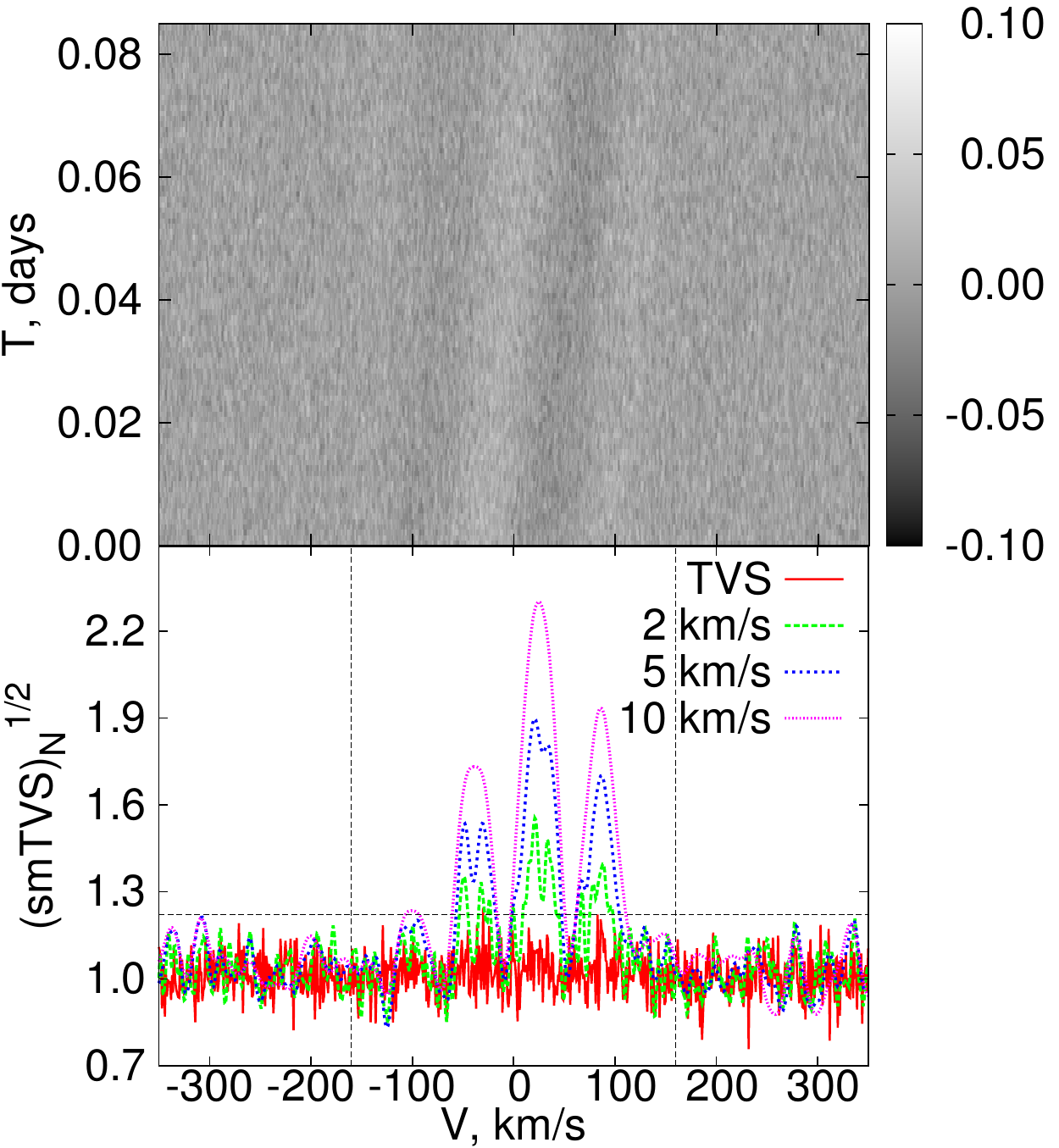}
  \caption{{\it Top panel:} The dynamic spectrum of 100 sequential model line profiles for NRP with parameters $\nu=3.4\,d^{-1}$, $l=5$, $m=-2$, $V\sin i=160\,$km/s, $W=5\,$km/s, $I_0=0.01$, $I_1=0.003$, $S/N=50$.  
{\it Bottom panel:} Normalized (smTVS)$_{\mathrm{N}}^{1/2}$ for different filter widths marked in the panel. The bottom solid red line is (TVS)$_{\mathrm{N}}^{1/2}$. The thin dashed line shows the significance level 0.001.The vertical dashed lines mark $\pm V\sin i$. }
  \label{Fig.NRP_SN50}
\end{figure}

To illustrate the efficiency of smTVS we will analyse variability related to NRP with parameters $\nu=3.4\,d^{-1}$, $l=5$, $m=-2$, $V\sin i=160\,$km/s, $W=5\,$km/s for low $S/N$ model spectra. The time interval between sequential profiles is $0.021^h$. The number of spectra is 100. The full length of simulated observations is $2.1^h$. The dynamic spectrum of model line profiles for low signal-to-noise ratio $S/N=50$ and the amplitudes of main frequency and its first harmonic  $I_0=0.01$, $I_1=0.003$ in continuum units respectively is presented in Fig.~\ref{Fig.NRP_SN50} (top).

For the residual profiles, showed in Fig.~\ref{Fig.NRP_SN50}~(top), smTVS were calculated for different filter widths $S$. The normalized (smTVS)$_{\mathrm{N}}^{1/2}$ for filter widths from $S=2\,$km/s to $S=10\,$km/s are given in Fig.~\ref{Fig.NRP_SN50}~(bottom panel). The lowest curve in the figure shows normalized (TVS)$_{\mathrm{N}}^{1/2}$.

For the adopted low $ S/N = 50 $ the pulsation amplitude is equal only 50\% of the average noise level. It is seen from the Fig.~\ref{Fig.NRP_SN50}, the contribution of the variable component of the profile associated with NRP in TVS does not stand out against the noise contribution. At the same time the smTVS clearly demonstrates the presence of weak LPV.

The amplitude of (smTVS)$_{\mathrm{N}}^{1/2}$ increases $\sim\sqrt{S}$ which let us detect even very weak variations of the line profile using suitable smoothing filter width $S$.

\subsection{TVS and NRP}

Firstly, we did TVS analysis for the model line profiles caused by NRP only. It turns out that the number of peaks $N_p$ in the TVS is related to $l$ and $m$ if the time sampling is uniform. 

For determination of $N_p$, only peaks whose amplitudes are higher then the amplitude of the noise in the TVS continuum region were counted. See, for example, Fig.~\ref{Fig.TVS_P} and Fig.~\ref{Fig.N_P}. The number of peaks in each panel of Fig.~\ref{Fig.TVS_P} denoted in Fig.~\ref{Fig.N_P} versus corresponding value of length of observations.

Unfortunately, there is no dependence between $N_p$ and pulsation mode for time sampling with gaps. If the uniform part is present in the data it can be used for the following analysis. The calculations and results described below are performed for uniform time sampling. 

\begin{figure}
\centering
  \includegraphics[width=0.8\columnwidth]{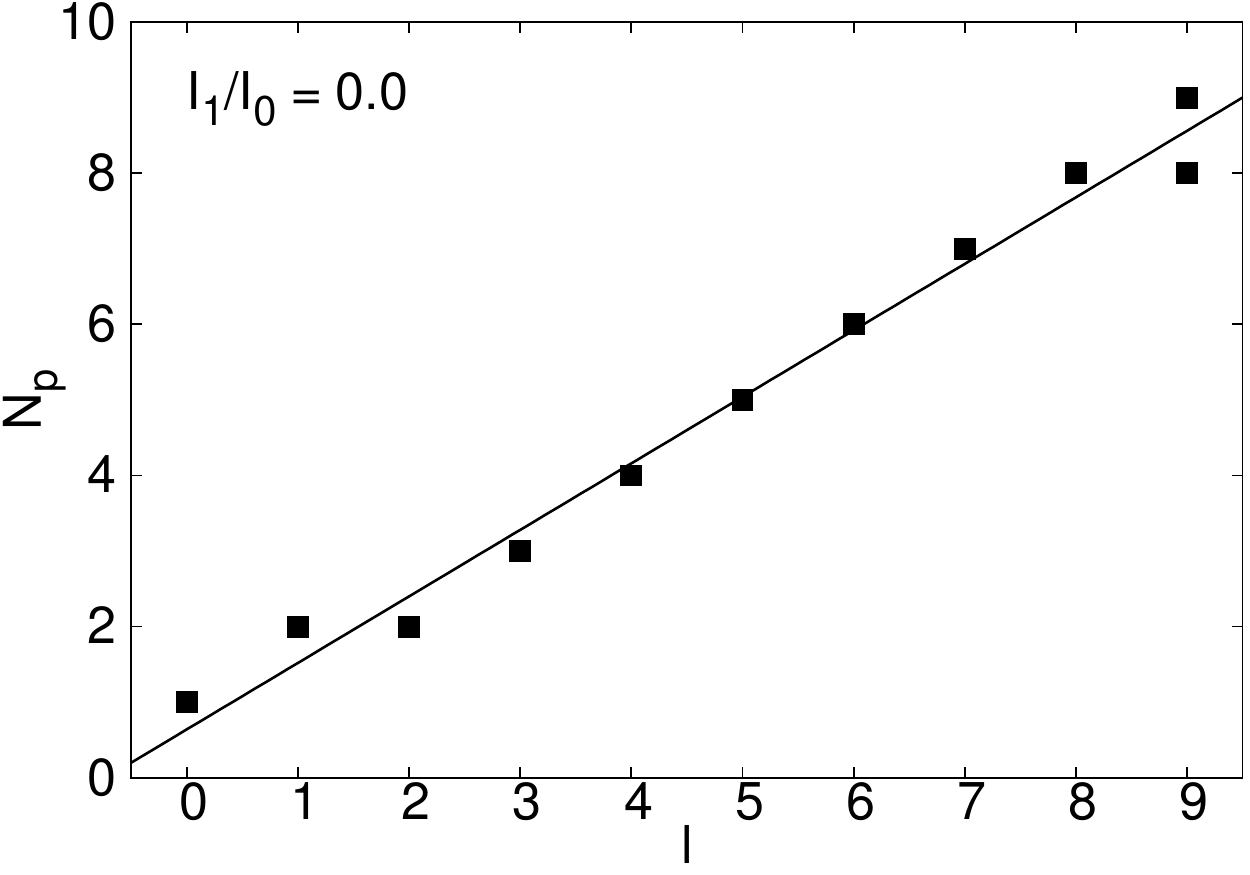}
  \caption{Number of peaks in (TVS)$_{\mathrm{N}}^{1/2}$ for different $l$ 
           obtained for 100 sequential spectra with parameters 
           ${I_1}/{I_0}=0$, $\nu=3.4\,$d$^{-1}$, $S/N=1000$. The solid line is least squares linear 
           approximation fit for this dependence.}
  \label{Fig.N_l}
\end{figure}

In a case if the length of observations ($T$) is less than $1/3$ of the pulsation period ($P$) and only main frequency ($I_0$) is presented or the ratio $I_1/I_0\leqslant 0.1$ the number of peaks in TVS equal to $l$ except $l=0$ (1 peak) and $l=1$ (2 peaks) (Fig.~\ref{Fig.N_l}). For $l=9$ the number of peaks can be $l-1$ because the amplitude of one peak can be small and in some random realizations it is very close to the noise level.

\begin{figure}
\centering
  \includegraphics[width=0.8\columnwidth]{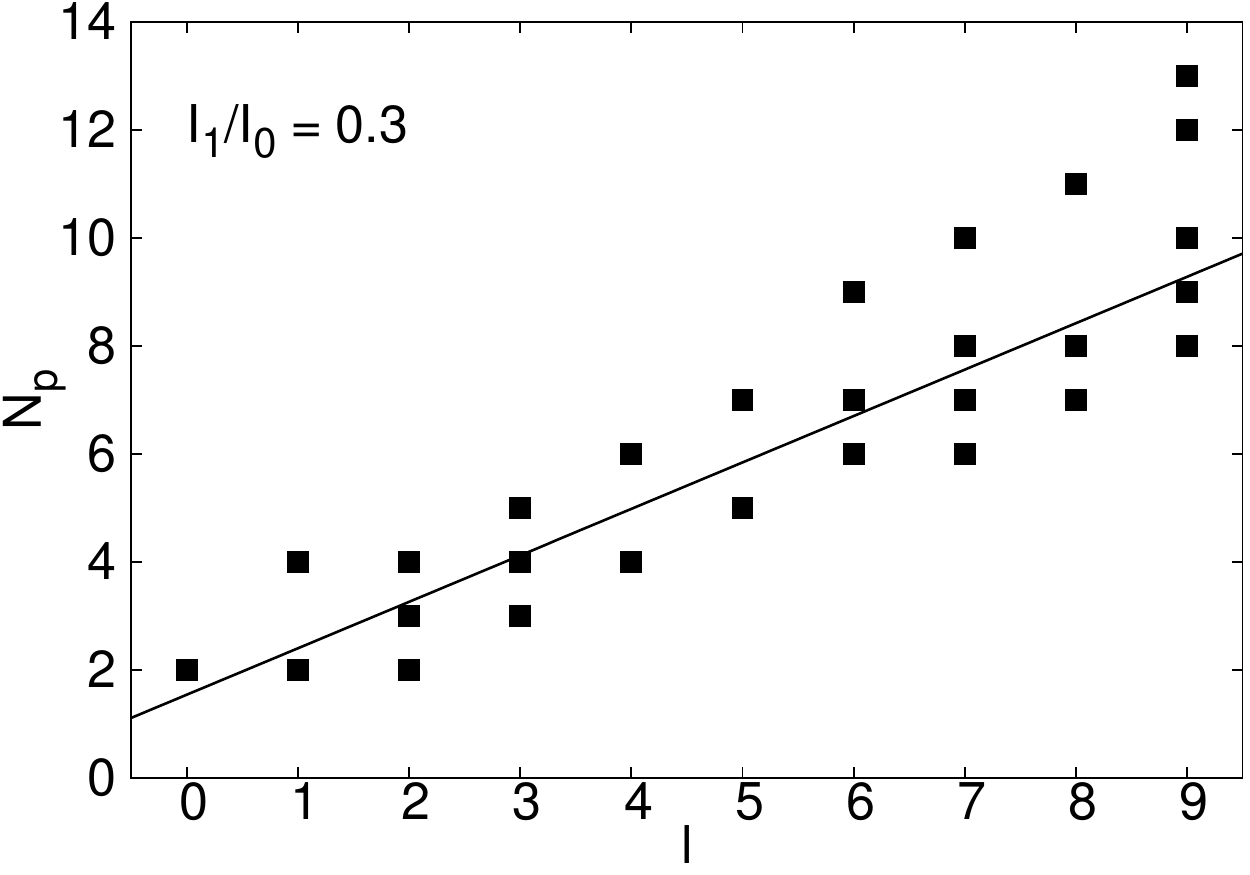}
  \caption{Number of peaks in (TVS)$_{\mathrm{N}}^{1/2}$ versus $l$ 
           obtained for 100 sequential spectra with parameters $I_1/I_0=0.3$,
           $\nu=3.4\,$d$^{-1}$, $S/N=1000$. The solid line is least squares linear fit.}
  \label{Fig.N_lm}
\end{figure}

\begin{figure}
\centering
  \includegraphics[width=0.8\columnwidth]{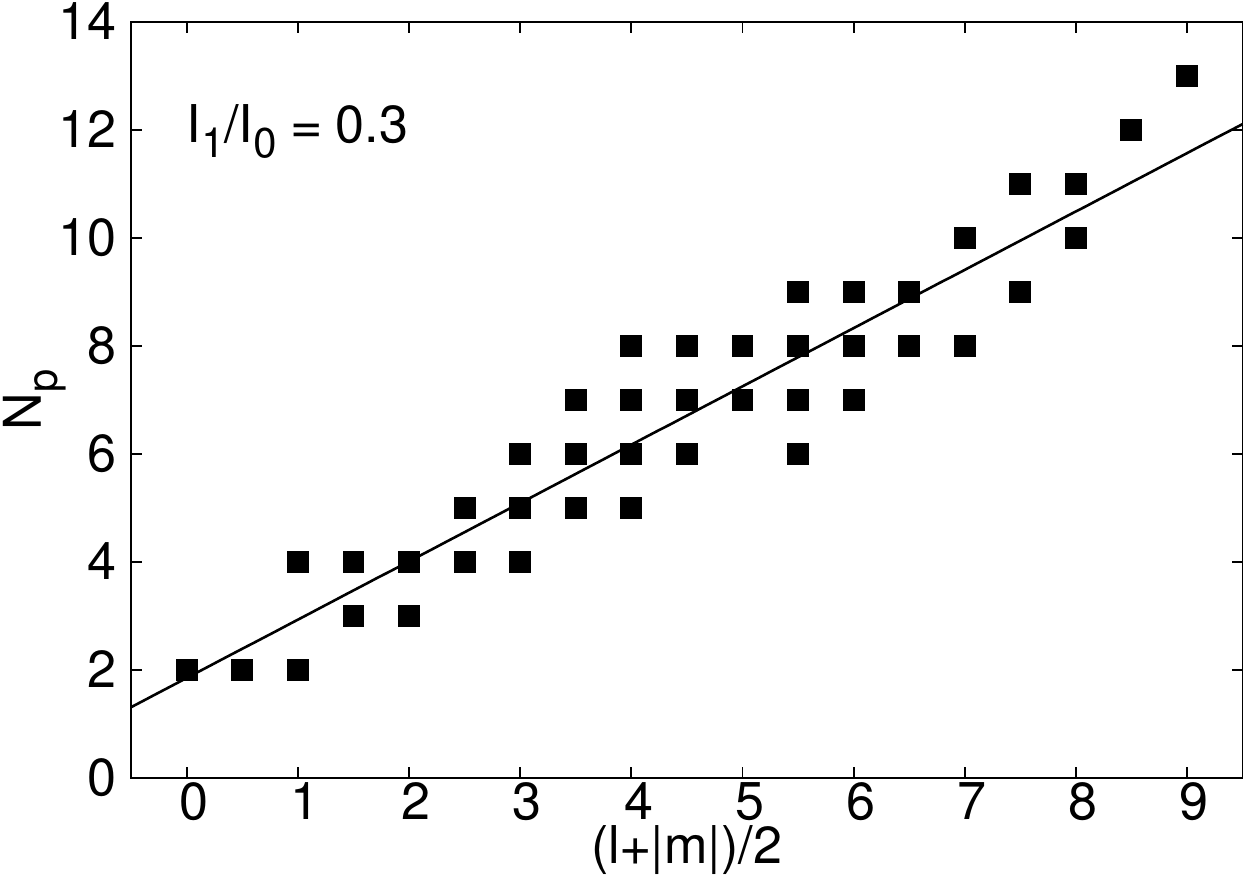}
  \caption{Number of peaks in (TVS)$_{\mathrm{N}}^{1/2}$ versus $(l+|m|)/2$ 
           obtained for 100 sequential spectra with parameters  
           $I_1/I_0=0.3$, $\nu=3.4\,$d$^{-1}$, $S/N=1000$. 
           The solid line is least squares linear fit.}
  \label{Fig.N_lm2}
\end{figure}

\begin{figure*}
  \includegraphics[width=2\columnwidth]{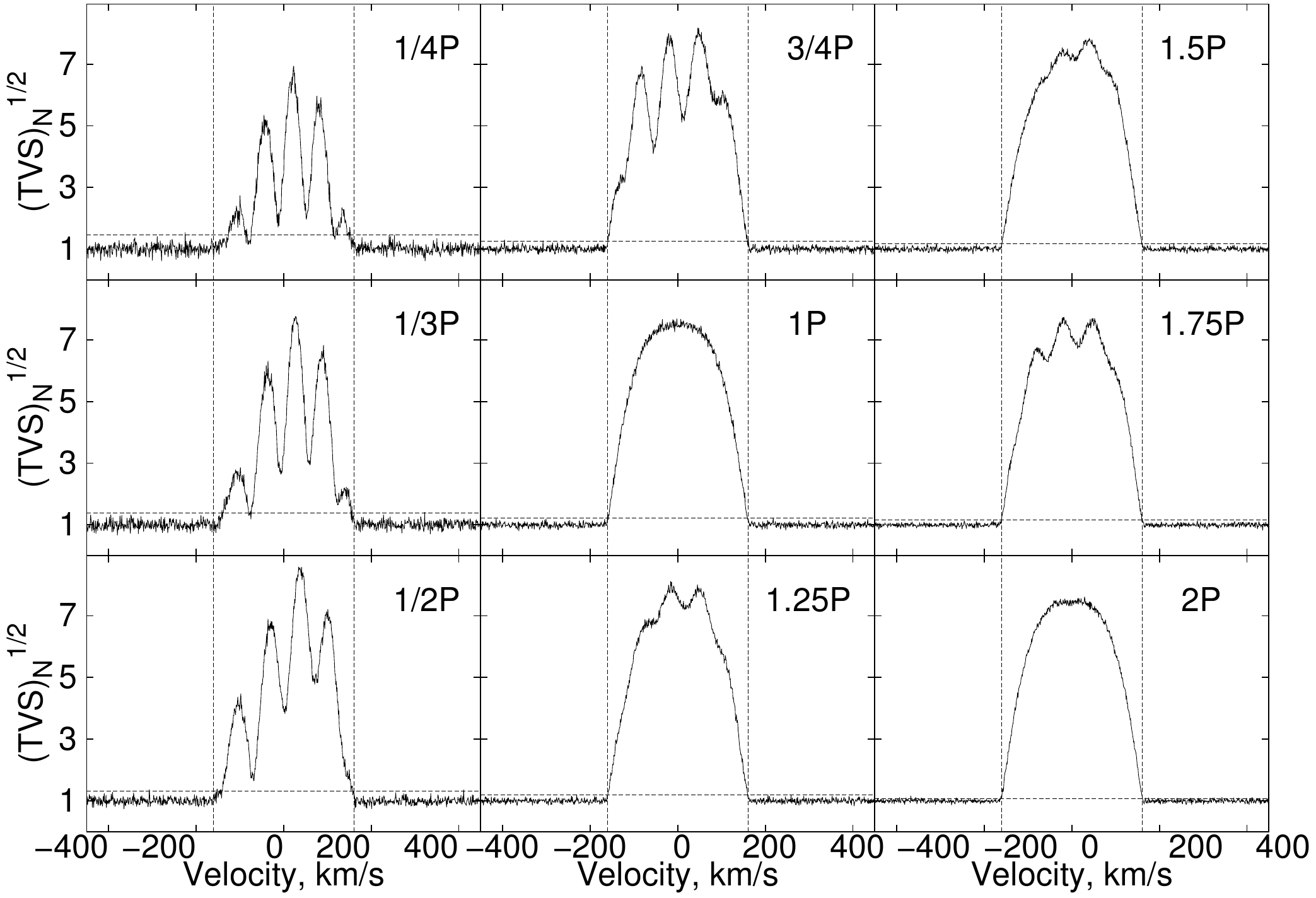}
  \caption{(TVS)$_{\mathrm{N}}^{1/2}$ for NRP in mode (5, -2) for different 
           length of observations in a fraction of pulsation period  ($P$) denoted on the panels,
           $\nu=3.4\,$d$^{-1}$,  $I_1/I_0=0.3$, $S/N=1000$. The number of spectra is 25 for the length of 
           observations equal 0.25P, 50 for 0.5P, 100 for P, 150 for 1.5P, 200 for 2P.
           The significance level 0.001 is shown by the dashed horizontal line. 
           The vertical dashed lines display $\pm V\sin i$.}
  \label{Fig.TVS_P}
\end{figure*}

\begin{figure}
\centering
  \includegraphics[width=0.8\columnwidth]{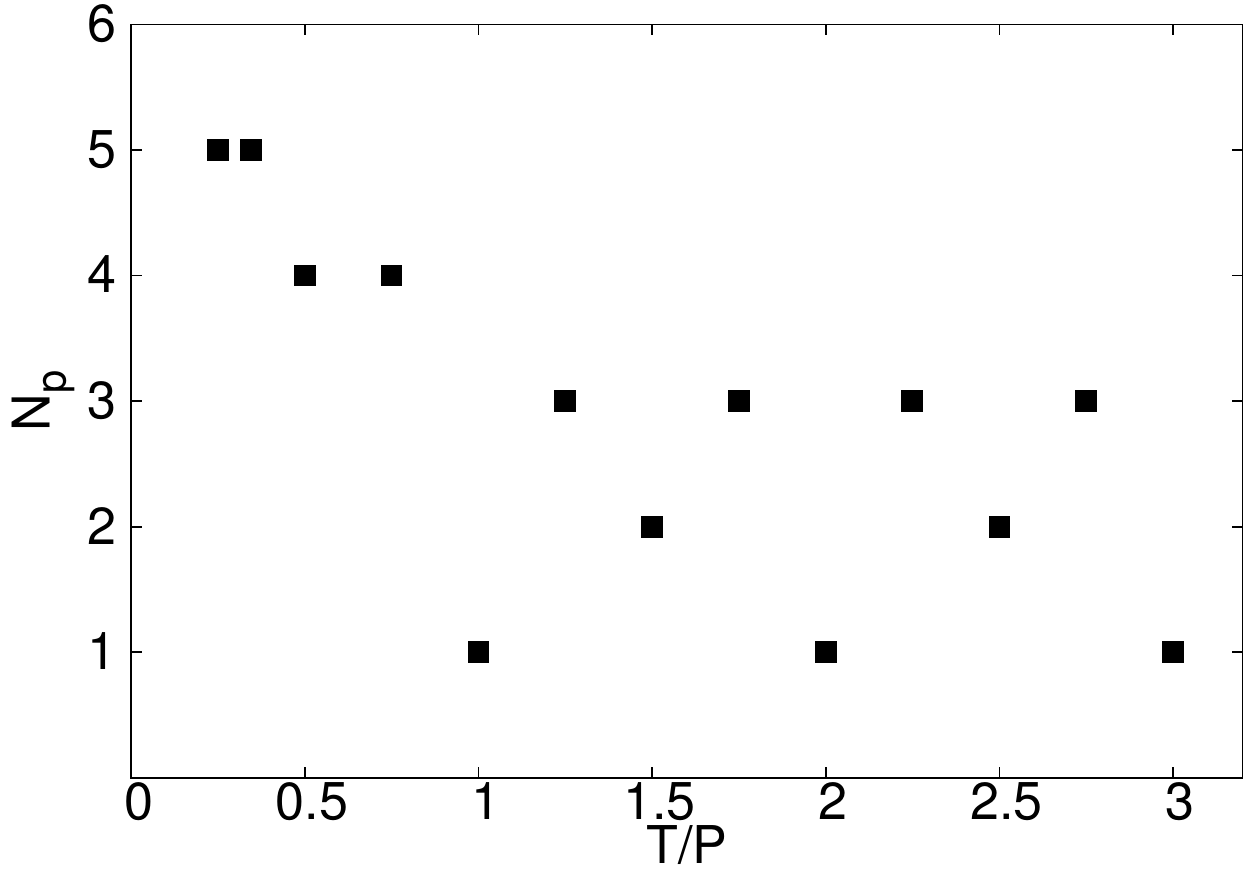}
  \caption{Number of peaks in (TVS)$_{\mathrm{N}}^{1/2}$ for different length of observations 
           as a function of time-series length (in units of pulsation period $P$).
           NRP mode (5, -2), $\nu=3.4\,$d$^{-1}$,  $I_1/I_0=0.3$, $S/N=1000$.
           The number of spectra is 25 for the length of 
           observations equal 0.25P, 50 for 0.5P, 100 for P, ..., 300 for 3P.}
  \label{Fig.N_P}
\end{figure}

When the ratio $I_1/I_0 > 0.1$ the number of peaks increases. Usually, the higher the value of $m$, the
higher the number of peaks. There is no clear dependence between $l$, $m$ and $N_p$ in this case. Fig.~\ref{Fig.N_lm} and Fig.~\ref{Fig.N_lm2} show $N_p$ in the TVS versus $l$ and $(l+|m|)/2$ respectively. Here for $l=9$ the number of peaks also can be $l-1$ like for $I_1/I_0\leqslant 0.1$ because the amplitude of the ninth peak is very close to the noise level. Note that the scattering of points in the figures stays small. 

Generally the number of peaks in TVS obeys the following conditions
\begin{equation}
\label{Eq.l_Npeaks}
l\leqslant N_P~~\mathrm{if}~~T\leqslant\frac{1}{3}P\,. 
\end{equation}

If $T>\frac{1}{3}P$ the number of peaks decreases and usually smaller than $l$. (TVS)$_{\mathrm{N}}^{1/2}$ for different length of observations in units of the pulsation period is shown in Fig.~\ref{Fig.TVS_P}. For $T=nP$, where $n=1,2,3...$ there is only one peak in the TVS. The number of peaks ($N_p$) in (TVS)$_{\mathrm{N}}^{1/2}$ versus length of observations (T) is shown in Fig~\ref{Fig.N_P}. This dependence can be useful to estimate the pulsation period. Denote via $t_k$ the time interval between first spectrum and the spectrum with number $k$. Then we can look for minima in the dependence $N_p(t_k)$. These minima correspond to values of $t_k = nP$.

\begin{figure}
  \includegraphics[width=1\columnwidth]{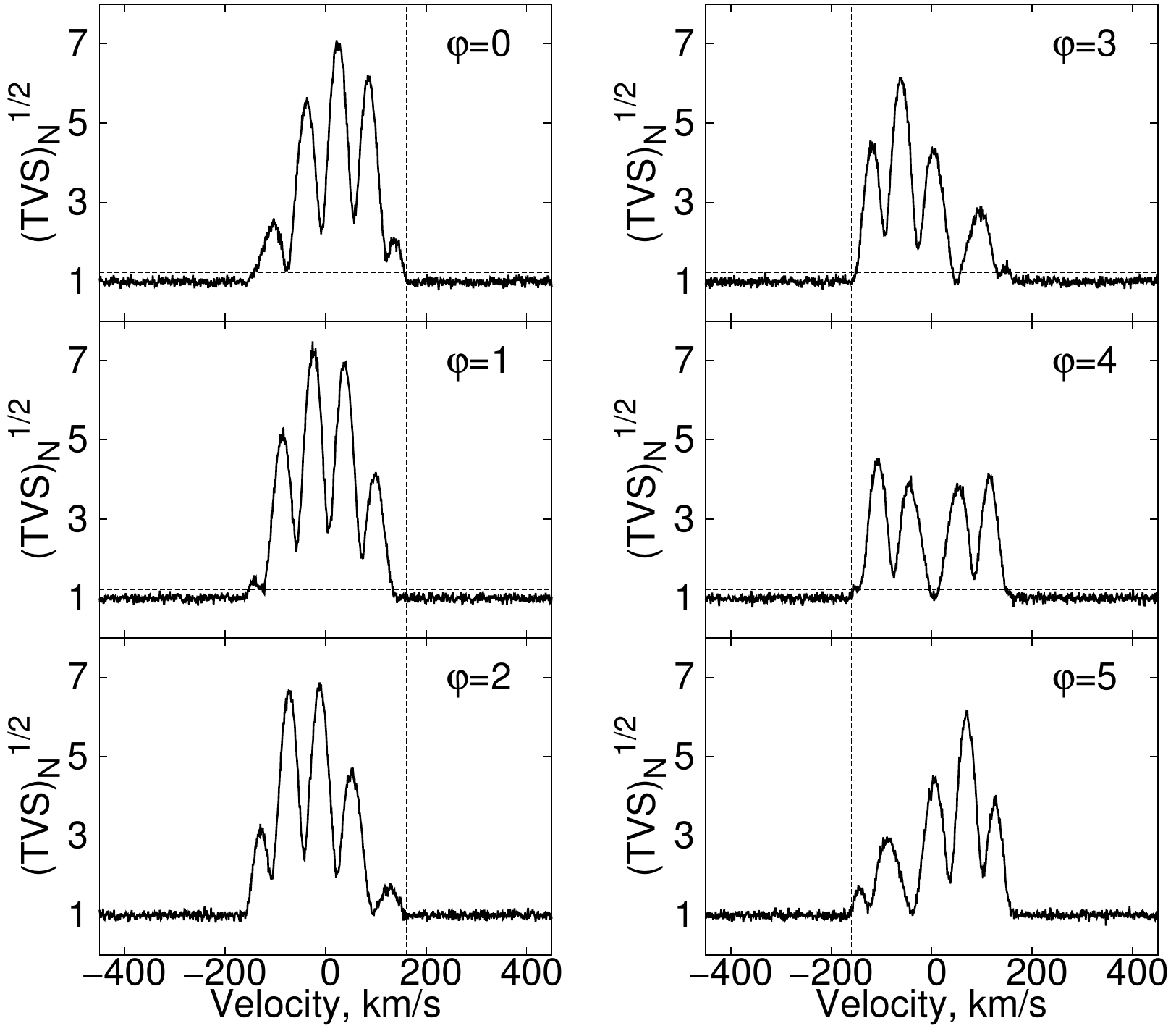}
  \caption{(TVS)$_{\mathrm{N}}^{1/2}$ for different initial phases. The phases are denoted on the panels.
           NRP mode is (5, -2), 100 sequential spectra, $\nu=3.4\,$d$^{-1}$,  $I_1/I_0=0.3$, $S/N=1000$.
           The significance level 0.001 is shown by 
           the dashed horizontal line. The vertical dashed lines show $\pm V\sin i$.}
  \label{Fig.Phase}
\end{figure}

The amplitude and the location of peaks in TVS also depends on the initial phase of NRP for the first spectrum from those which were used to calculate the TVS. The number of peaks stays the same for most values of the initial phase but it can change for few of them. In Fig.~\ref{Fig.Phase} the (TVS)$_{\mathrm{N}}^{1/2}$ for pulsation mode (5, -2) for different initial phases is shown. Although the TVS for different initial phases look different it is difficult to estimate the initial phase from the TVS without additional modelling.

\begin{figure*}
  \includegraphics[width=2\columnwidth]{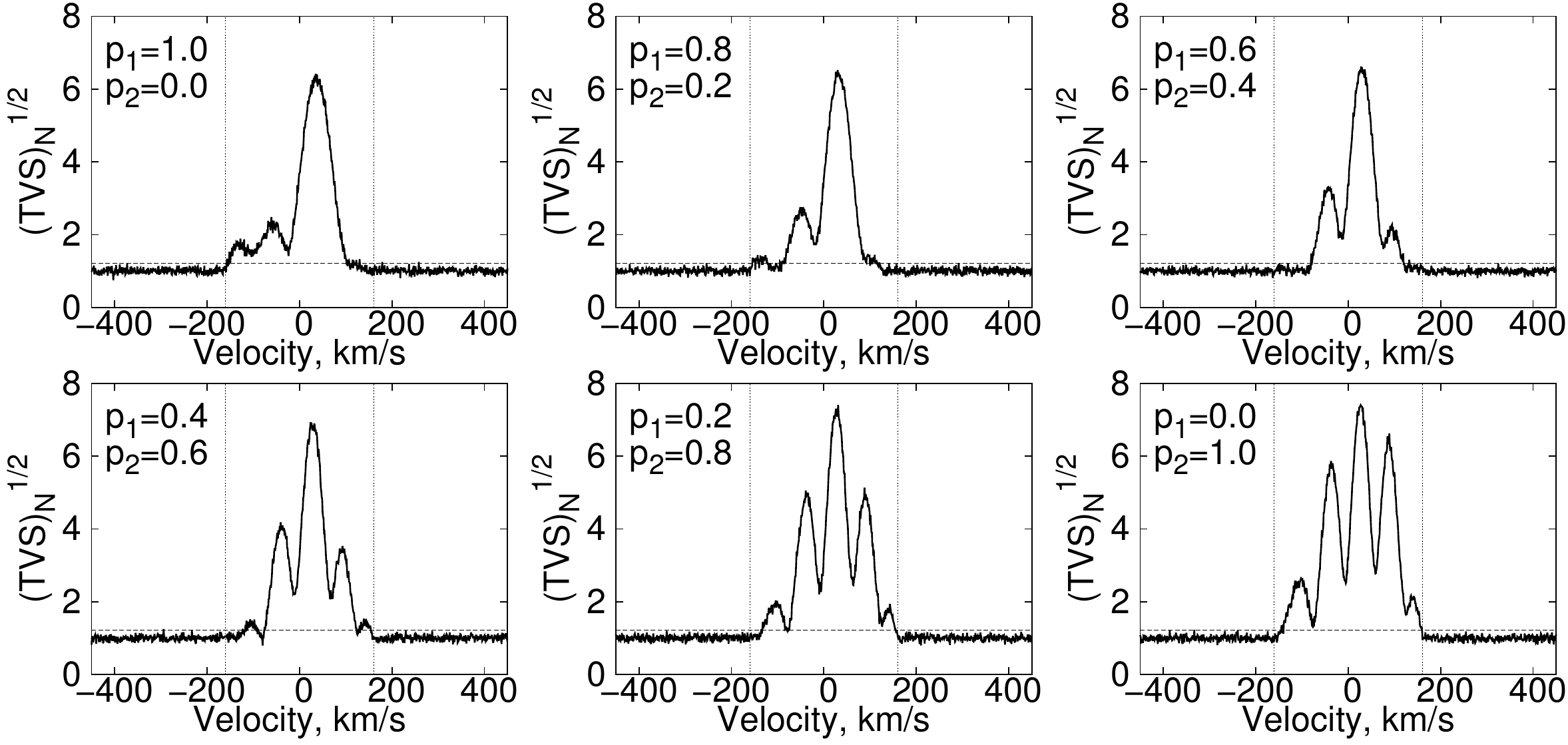}
  \caption{(TVS)$_{\mathrm{N}}^{1/2}$ obtained for 100 sequential spectra for NRP in two modes $(l,m)=(2,-1)$, 
           $\nu=3.4\,$d$^{-1}$ and $(l,m)=(5,-2)$, $\nu=4.6\,$d$^{-1}$,  $I_1/I_0=0.3$, $S/N=1000$,
           $T=1.7^h$ ($T\leqslant\frac{1}{3}P$, where P is the pulsation period). 
           Here $A_1+A_2=0.01=const$, where $p_1=\frac{A_1}{A_1+A_2}$, 
           $p_2=\frac{A_2}{A_1+A_2}$, $A_1$ and $A_2$ 
           are the amplitudes of pulsation in modes $(l,m)=(2,-1)$ and $(l,m)=(5,-2)$
           respectively. The vertical dashed lines show $\pm V\sin i$. The horizontal dashed line
           is the significance level 0.001.}
  \label{Fig.TVS_2l}
\end{figure*}

We also modelled TVS for multiple NRP modes. The (TVS)$_{\mathrm{N}}^{1/2}$ for two pulsation modes $(l,m)=(2,-1)$,  $\nu=3.4\,$d$^{-1}$ and $(l,m)=(5,-2)$, $\nu=4.6\,$d$^{-1}$ with different amplitudes and $T\leqslant\frac{1}{3}P$ obtained for 100 sequential spectra is shown in Fig.~\ref{Fig.TVS_2l}. In the figure $A_1+A_2=0.01=const$, where $p_1=\frac{A_1}{A_1+A_2}$, $p_2=\frac{A_2}{A_1+A_2}$, $A_1$ and $A_2$ are the amplitudes of pulsations in modes $(l,m)=(2,-1)$ and $(l,m)=(5,-2)$ respectively.
In the case of multiple NRP modes equation~(\ref{Eq.l_Npeaks}) is valid at least for lowest $l$. 

For low quality data equation~(\ref{Eq.l_Npeaks}) should be used with caution. The number and the amplitude of
peaks in the TVS also depend on the amplitude of pulsations, the S/N ratio and the number of spectra. Depending on the combination of these three factors there could be one or several peaks in the TVS (smTVS). The smaller amplitude of pulsations the smaller the amplitude of the peaks. The lower the S/N and the fewer number of spectra the fewer number of peaks can be seen. For high modes decreasing of $N_p$ goes faster with decreasing S/N, number of spectra, and pulsation amplitude than for low modes. Using smTVS can add some extra peaks if the filter width is small. For higher filter width some small amplitude peaks can disappear but the high amplitude peaks remain visible.

To estimate the pulsation mode using smTVS one has to play with the smoothing filter width. A good
strategy is to start with a very small filter width and to increase it. Firstly, the number of peaks in smTVS will increase. The amplitude of peaks will increase. For some value of the filter width, the number of peaks will reach its maximum and when increasing the filter width further the number of peaks will decrease. This happens because by increasing the filter width, the amplitude and the width of the peaks increase and eventually the peaks will merge. The smTVS for the filter width with maximum number of peaks should be used to estimate the pulsation mode. Note, that this method gives only estimation of the pulsation mode. 

When a different filter width changes a number of peaks, smTVS can be used to estimate the
pulsation mode. It is also possible that for some value of $S$ the smTVS will show extra peaks but the
number of peaks will be fewer than it will be for good quality data. If the number of peaks is one for any filter width the smTVS  cannot be used for pulsation mode estimation. If the amplitude of pulsation is small (for example 0.001) even for high quality data smTVS will detect variability but will show only one peak for different filter widths. In this case the smTVS cannot be used for the pulsation mode estimation.

\subsection{NRP and Rotational modulation}

To analyse the contribution of rotation on the shape of TVS we performed the TVS analysis for the spectra affected by rotational modulation only. We calculated the spectra according to~\citet{HenrichsSudnik2014} model where the line profile variability induced by the stellar prominences represented as a spheres touching the stellar surface. The prominence will give extra absorption or emission in the line profile depending on its location in the line of sight. After that we computed spectra caused by rotational modulation and NRP both. Then the TVS for the spectra affected by rotational modulation only, by NRP only, and by NRP and rotational modulations both was calculated.

\begin{figure}
\centering
  \includegraphics[width=0.8\columnwidth]{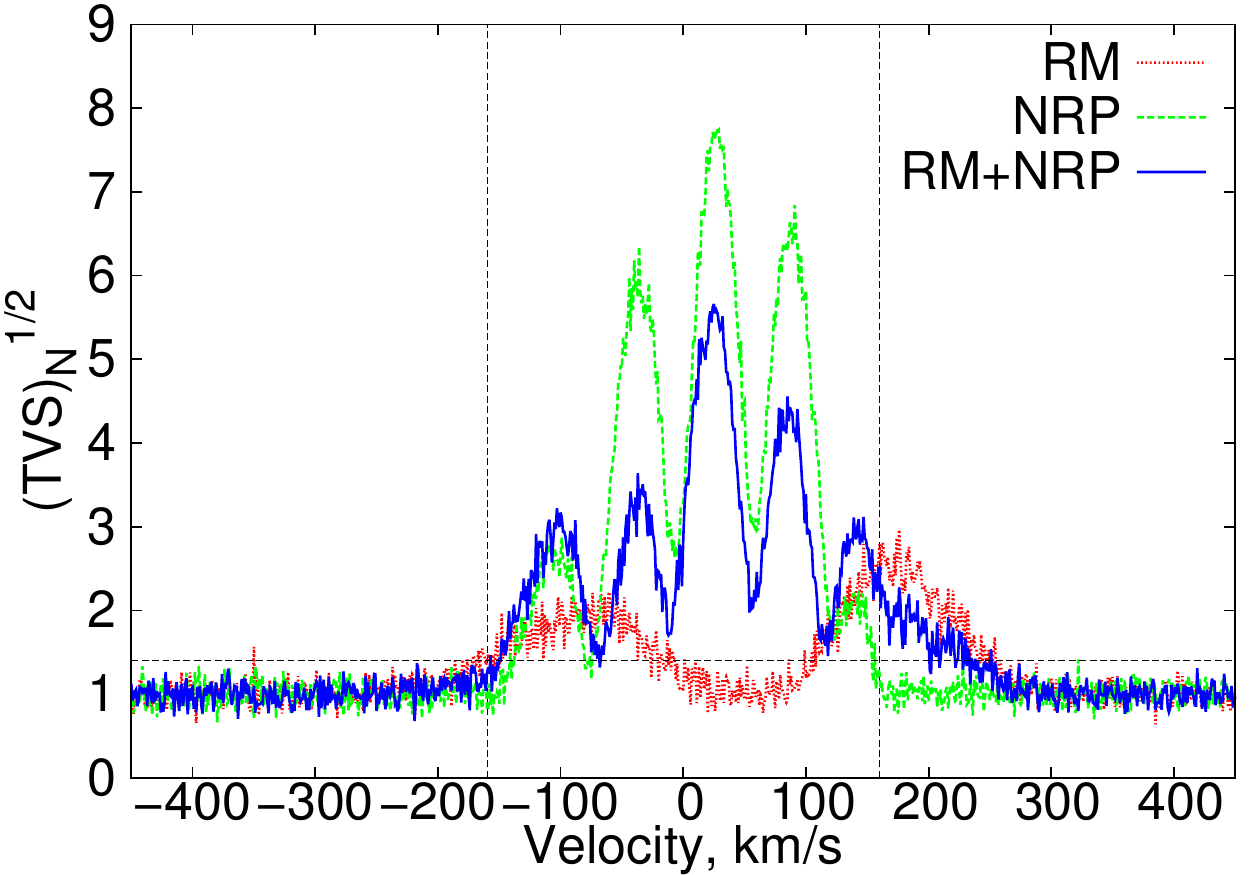}
  \caption{(TVS)$_{\mathrm{N}}^{1/2}$ for rotational modulation caused by 
           five prominences randomly located around the star (red dotted line), by
           NRP in the mode (5,-2) (green dashed line) and
           for their superposition (blue solid line).
           The length of simulated observations is 2.1$^h$, 
           the number of spectra included in time series is 33, the pulsation period is 7.06$^h$, 
           the rotation period is 4$^d$, $V \sin i=160\,$km/s (marked by the vertical dashed lines), 
           and $S/N=1000$.}
  \label{Fig.TVS_RMNRP2}
\end{figure}

Fig.~\ref{Fig.TVS_RMNRP2} shows the (TVS)$_{\mathrm{N}}^{1/2}$ for 5 prominences randomly located around the star. The configuration of the prominences does not change with time. Here the length of simulated observations ($T$) is 2.1$^h$, the number of spectra included in the time series is 33, the pulsation period is 7.06$^h$, the pulsation mode (5,-2), the rotation period is 4$^d$, $V \sin i=160\,$km/s, $S/N=1000$. The amplitude of the TVS for rotational modulation (red dotted line) is small. The rotation modulation does not change the number of peaks in the TVS for superposition of rotational modulation and NRP (solid blue line). Though the amplitude of the TVS for superposition of rotational modulation and NRP  is smaller that for NRP only (green dashed line) the number of peaks stays the same. In this case it is possible to estimate mode $l$ using TVS.

\begin{figure}
\centering
  \includegraphics[width=0.8\columnwidth]{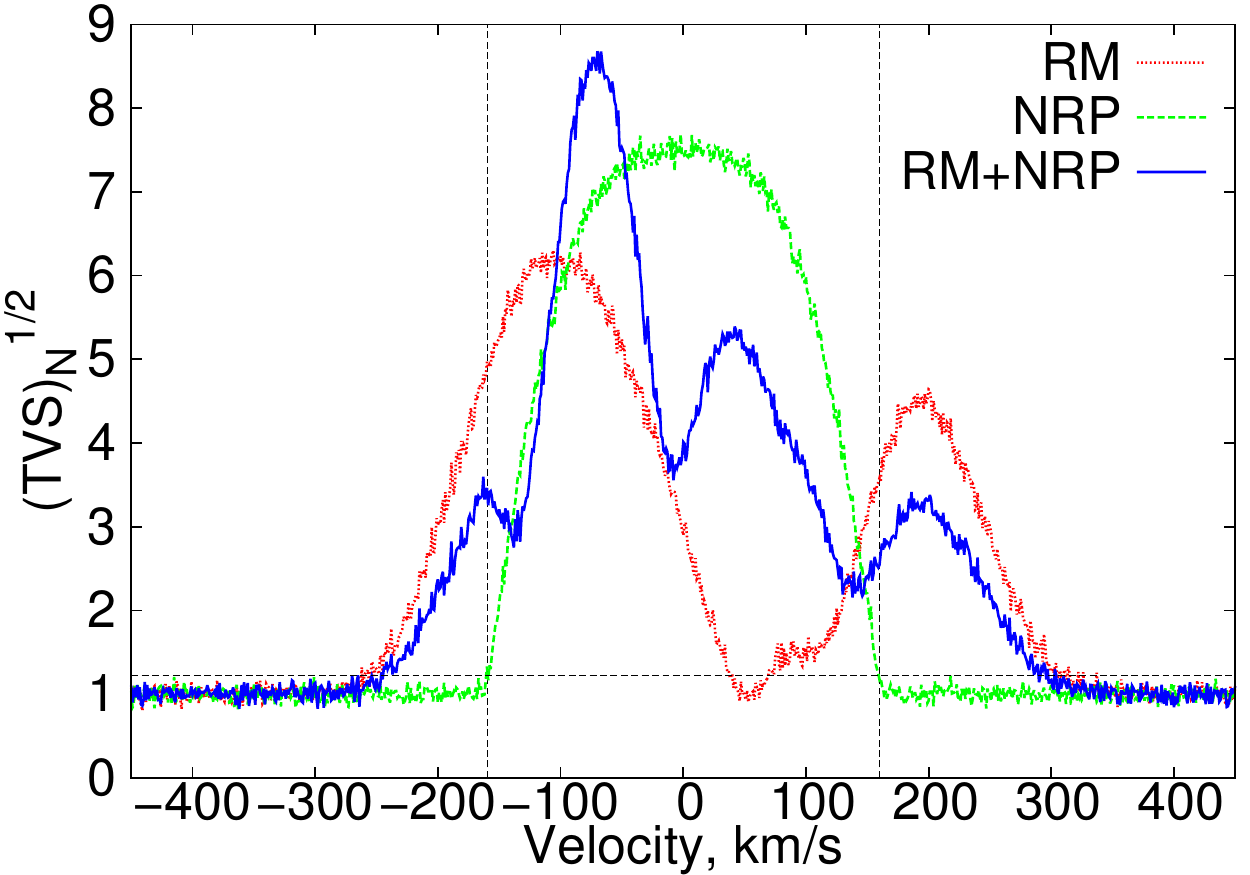}
  \caption{The same as in Fig.~\ref{Fig.TVS_RMNRP2} but
           the length of simulated observations is 7.06$^h$, the number of spectra is 100.}
  \label{Fig.TVS_RMNRP8}
\end{figure}

\begin{figure}
\centering
  \includegraphics[width=0.8\columnwidth]{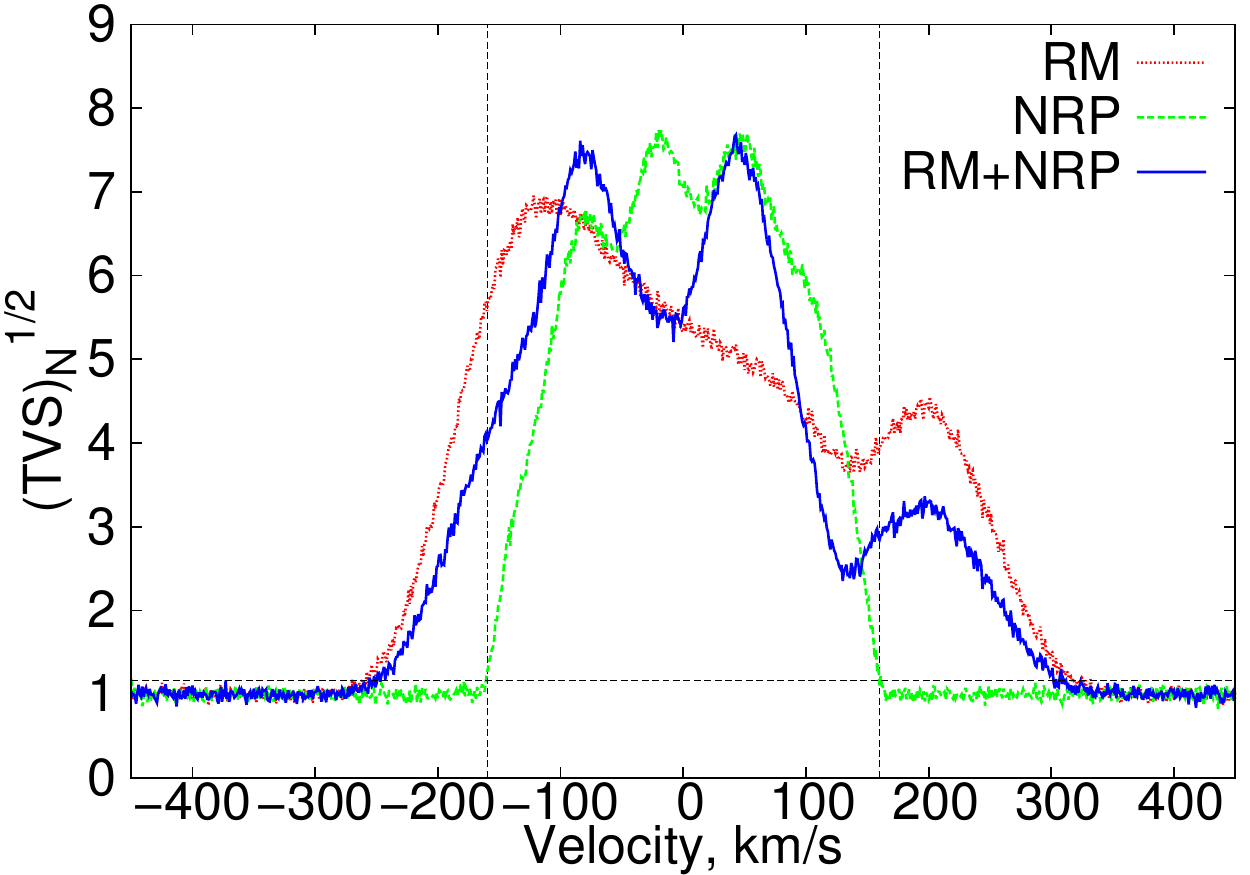}
  \caption{The same in as Fig.~\ref{Fig.TVS_RMNRP2} but
           the length of simulated observations is 12.35$^h$, the number of spectra is 175.}
  \label{Fig.TVS_RMNRP14}
\end{figure}

For longer $T$ the impact of rotation on the shape of resulting TVS increases. The amplitude of TVS related with rotation increases and the shape of TVS caused by superposition of NRP and rotational modulation would be different from both of them separately (Fig.~\ref{Fig.TVS_RMNRP8} and Fig.~\ref{Fig.TVS_RMNRP14}). In this situation the TVS (smTVS) cannot be used for the estimation of the NRP modes and the pulsation period.

\subsection{Interpretation of the observations}

Weak variability of the lines of the ions Si, C, O, N, Mg in the spectra of programme O stars revealed by smTVS only was reported in our previous papers \citep{Kholtygin2003, Kholtygin2006, Kholtygin2011, Sudnik2012}.

To estimate pulsation modes the TVS (smTVS) was computed for datasets that have eight or more spectra per 1--3$^h$  observations. This time interval was chosen taking into account that equation~(\ref{Eq.l_Npeaks}) is valid if $T\leqslant\frac{1}{3}P$. It is not reasonable to consider a smaller number of spectra during 1 to 3$^h$ observations.

Our source list yields only two datasets that are suitable for estimation NRP mode using TVS (smTVS). The spectra were divided into three groups with the length of observations approximately one hour, two hours and three hours. For these groups both TVS and smTVS were calculated. 

\begin{figure*}
  \includegraphics[width=2\columnwidth]{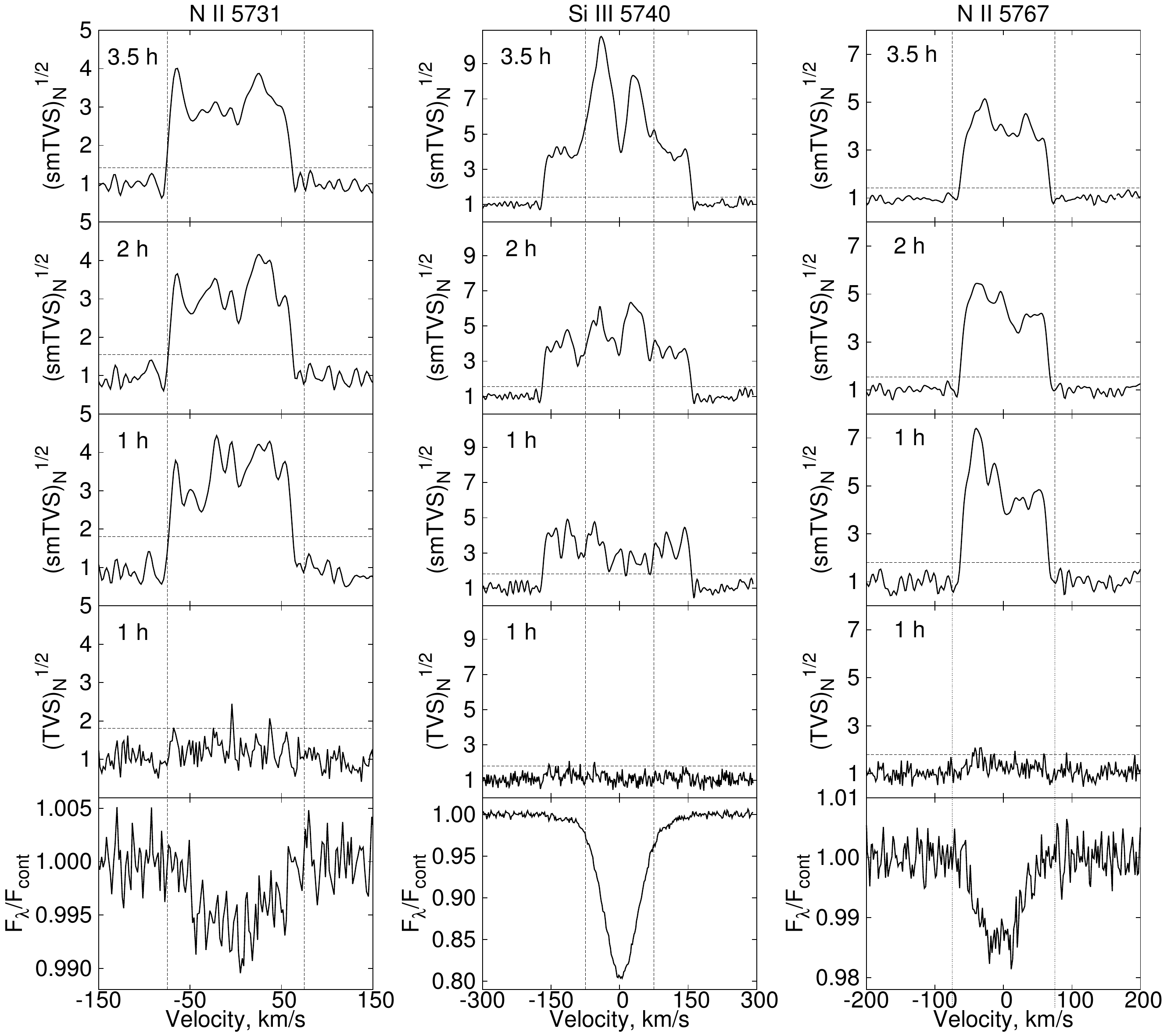}
  \caption{The N\,II~$\lambda\,5731\,$\AA{}, Si\,III~$\lambda\,5740\,$\AA{}, 
           N\,II~$\lambda\,5767\,$\AA{} lines in the spectra of $\rho\,$Leo.
           From bottom to top: the line profile, the (TVS)$_{\mathrm{N}}^{1/2}$ for 
           the length of observation 0.98$^h$,  (smTVS)$_{\mathrm{N}}^{1/2}$ for 
           the length of observation 0.98$^h$, (smTVS)$_{\mathrm{N}}^{1/2}$ for 
           the length of observation 2.01$^h$, (smTVS)$_{\mathrm{N}}^{1/2}$ for 
           the length of observation 3.54$^h$.  The filter width $S$ is 0.1\,\AA.
           The horizontal line corresponds to the significance level 0.001. 
           The vertical dashed lines denote $\pm V \sin i$.}
  \label{Fig.TVS_roLeo}
\end{figure*}

The (TVS)$_{\mathrm{N}}^{1/2}$ and (smTVS)$_{\mathrm{N}}^{1/2}$ for the N\,II~$\lambda\,5731\,$\AA{}, Si\,III~$\lambda\,5740\,$\AA{}, N\,II~$\lambda\,5767\,$\AA{} lines profiles in the spectra of star $\rho\,$Leo are shown in Fig.~\ref{Fig.TVS_roLeo}. The star is slow rotator ($V\sin i$=75\,km/s). The rotation would not substantially affect the TVS (smTVS) for the length of observations 3.5$^h$ and less.

The TVS obtained for the length of observations 1$^h$ does not show variability for all three lines. In the same way the smTVS for the same length of observations with the filter width $S=0.1\,$\AA\ demonstrates the existence of LPV at the significance level $\alpha=10^{-3}$. 

The number of peaks in smTVS is different for these three lines and decreases with increasing length
of observations. There are four peaks in smTVS for the line N\,II~$\lambda\,5767\,$\AA{} (right panel) 
and $T=1^h$. The number of peaks is reduced to three for $T=2^h$ and to two for $T=3.5^h$. The number of peaks for the line Si\,III~$\lambda\,5740\,$\AA{} (middle panel) in the $V\sin i$ band decreases from 4 for $T=1^h$ 
to 2 for $T=3.5^h$. The number of peaks for the line N\,II~$\lambda\,5731\,$\AA{} (left panel) (3--4) 
practically does not change for all lengths of observations. The decrease of the number of peaks in the smTVS with increasing length of observation is consistent with our model. To search the pulsation period we need 
a longer dataset as long as rotation would not influence TVS for $T=P$.

To estimate the pulsation mode $l$ we can use the number of peaks in smTVS for minimal length of observations $T=1^h$. The smTVS shows different number of peaks for the N\,II~$\lambda\,5731\,$\AA{}, Si\,III~$\lambda\,5740\,$\AA{}, N\,II~$\lambda\,5767\,$\AA{} lines. The mode $l$ would be equal or less than minimum number of peaks. Thus $l\le 4$. Using the linear fit in Fig.~$\ref{Fig.N_lm}$ we can estimate $l=3-4$. This value is agree with an estimation $l\ge 2$ by ~\citet{Kholtygin2007}.  

The TVS and smTVS for the line Si\,III~$\lambda\,5740\,$\AA{} present special interest. We see in the smTVS for this line in Fig.~\ref{Fig.TVS_roLeo} (middle panel) (unlike the smTVS for the N\,II~$\lambda\,5731\,$\AA{} and N\,II~$\lambda\,5767\,$\AA{} lines) the strong contribution in the region outside the $\pm V\sin i$ range. Furthermore, the amplitude of the smTVS beyond the  $\pm V\sin i$ range is nearly constant in contrast with the smTVS inside the $\pm V\sin i$ range. Moreover, the number of small peaks in the parts of smTVS to the left and right of $\pm V\sin i$ range is close to the number of peaks inside this range. 

The cause of such behaviour is not clear. We can conclude that it may be connected with the modulation 
of the stellar wind by the NRP. This possibility was noted by \citet{Kholtygin2006}. 
When such modulation is extended far from the stellar surface the effect of absorption due to the
matter of the stellar wind in the line of sight between the star and the observer leads to repeating 
the shape of both the TVS and the smTVS inside and outside the $\pm V\sin i$ band.
 
\begin{figure*}
  \includegraphics[width=2\columnwidth]{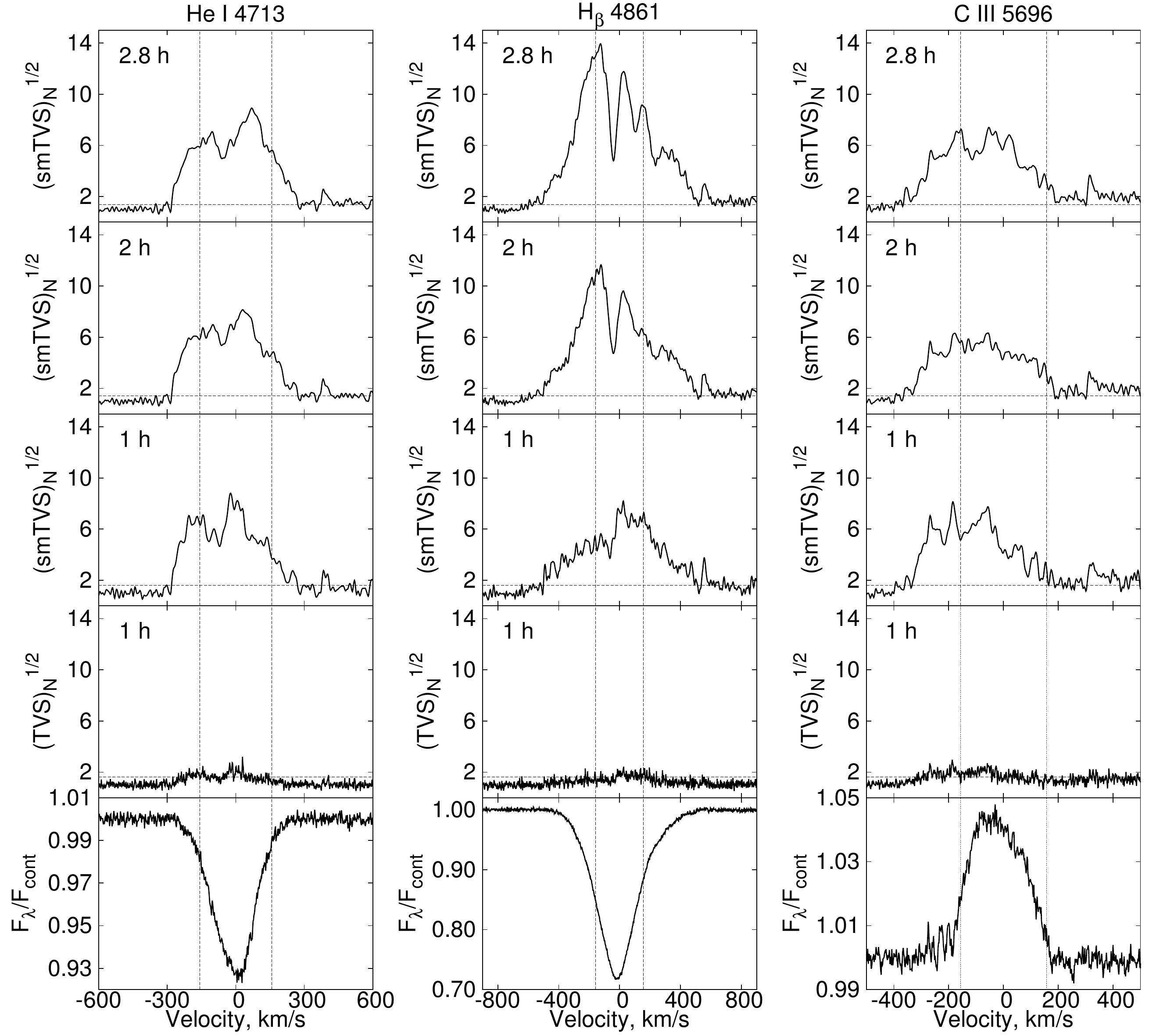}
  \caption{The He\,I~$\lambda\,4713\,$\AA{}, H$_\beta$~$\lambda\,4861\,$\AA{}, 
           C\,III~$\lambda\,5696\,$\AA{} lines in the spectra of $\delta\,$Ori~A. 
           From bottom to top: the line profile, the (TVS)$_{\mathrm{N}}^{1/2}$ for 
           the length of observation 1.01$^h$,  (smTVS)$_{\mathrm{N}}^{1/2}$ for 
           the length of observation 1.01$^h$, (smTVS)$_{\mathrm{N}}^{1/2}$ for 
           the length of observation 2.02$^h$, (smTVS)$_{\mathrm{N}}^{1/2}$ for 
           the length of observation 2.8$^h$.  The filter width $S$ is 0.1\,\AA. 
           The horizontal line corresponds to the significance level 0.001. 
           The vertical dashed lines show $\pm V \sin i$.}
  \label{Fig.TVS_dOri}
\end{figure*}

The TVS for $\delta\,$Ori~A (Fig.~\ref{Fig.TVS_dOri}) for $T=1^h$ shows the LPV but the amplitude of the TVS is small and does almost not exceed the significance level $\alpha=10^{-3}$. The smTVS for $T=1^h$ indicates that smTVS is definitely above this level and the peaks in smTVS are clearly seen. 

The number of peaks in smTVS for the C\,III~$\lambda\,5696\,$\AA{} line (right panel) in the $\pm V\sin i$ 
range is 3 for the length of observations $T=1^h$ and  $T=3.5^h$. For $T=2^h$ the number of peaks 
is difficult to estimate. Such behaviour of smTVS for the C\,III~$\lambda\,5696\,$\AA{} line can be explained by the modulation of the stellar wind (see above). The shape of smTVS beyond the $\pm V\sin i$ range is strongly asymmetric. This asymmetry can be connected with the contribution of the component Ab of $\delta\,$Ori~A in NRP (see, for example, \citealt{Kholtygin2006}).

The number of peaks in the He\,I~$\lambda\,4713\,$\AA{} line (left panel) is two and does not strongly change by increasing the length of the observations. In smTVS for the line H$_\beta$ (middle panel) the number of peaks also does not exceed 2 in the $\pm V\sin i$ range. So, we can conclude that the mode $l\le 2$ in line with the conclusion by \citet{Kholtygin2006} that the main component Aa$^1$ of the triple system $\delta\,$Ori~A pulsates in the NRP (l,m) mode (2,-2).

The different number of peaks observed in different lines can, in principle, be connected with various values of the intrinsic line width $W$. More over, these lines can be formed at the various depth in the stellar atmosphere and hence the velocity field in the line formation area could be different. 

The rotation modulation does almost not change the number of peaks for such length of observations but it can change the amplitude and the shape of peaks in TVS (smTVS) as it was shown in Sect. 6.3. If some lines are more sensitive to local inhomogeneities in the stellar atmosphere than others then rotation modulation will have various  influence on different lines profiles. Consequently the amplitude and the shape of peaks in TVS (smTVS) will be different for different lines.

\section{Conclusions}
We demonstrated that the method of smTVS analysis is very effective in detecting micro LPV that
cannot be discovered the usual way. This method can be used when amplitude of the profile variations is small (less than 1\% in continuum unit) and does not exceed the noise level, the number of spectra is small and the time series are irregular.

Weak variability of the lines of the ions N, Mg, S in spectra of stars $\rho\,$Leo
and $\lambda\,$Cep was revealed by smTVS only.

We show how the TVS and smTVS analysis of the spectra of non-radially pulsating stars can be used for the 
estimation of a pulsation mode $l$.

\section*{Acknowledgments}
We would like to acknowledge our anonymous referee for very useful comments and suggestions which improved an earlier  version of the manuscript. We acknowledge Saint-Petersburg State University (SPbGU) for the research grants 6.50.1555.2013 and 6.38.18.2014.

\label{lastpage}

\end{document}